\documentclass[aps,prc,preprint,superscriptaddress,floatfix,nofootinbib]{revtex4-1} 
\usepackage{amsmath}
\usepackage{mathtools}
\usepackage[english]{babel}
\usepackage[bookmarks=false,colorlinks=false]{hyperref}
\usepackage{graphicx}
\usepackage{url}
 \usepackage{footnote}
 \newcommand{\eq}[1]{\begin{align} #1 \end{align}}
 
\begin{document}

\title{Bridging the gap between event-by-event fluctuation
  measurements and theory predictions in relativistic nuclear
  collisions}

\author{P.~Braun-Munzinger}
\affiliation{Extreme Matter Institute EMMI, GSI, Darmstadt, Germany}
\affiliation{Physikalisches Institut, Universit\"{a}t Heidelberg, Heidelberg, Germany}
\author{A.~Rustamov}
\affiliation{Physikalisches Institut, Universit\"{a}t Heidelberg, Heidelberg, Germany}
\affiliation{National Nuclear Research Center, Baku, Azerbaijan}
\author{J.~Stachel}
\affiliation{Physikalisches Institut, Universit\"{a}t Heidelberg, Heidelberg, Germany}

\begin{abstract}
We develop methods to deal with non-dynamical contributions to
event-by-event fluctuation measurements of net-particle numbers in
relativistic nuclear collisions. These contributions arise from impact
parameter fluctuations and from the requirement of overall net-baryon
number or net-charge conservation and may mask the dynamical
fluctuations of interest, such as those due to critical endpoints in
the QCD phase diagram. Within a model of independent particle sources
we derive formulae for net-particle fluctuations and develop a
rigorous approach to take into account contributions from participant
fluctuations in realistic experimental environments and at any
cumulant order. Interestingly, contributions from participant
fluctuations to the second and third cumulants of net-baryon
distributions are found to vanish at mid-rapidity for LHC energies
while higher cumulants of even order are non-zero even when the net-baryon 
number at mid-rapidity is zero. At lower beam energies the
effect of participant fluctuations increases and induces spurious
higher moments. The necessary corrections become large and need to be
carefully taken into account before comparison to theory. We also
provide a procedure for selecting the optimal phase-space coverage of
particles for fluctuation analyses and discuss quantitatively the
necessary correction due to global charge conservation.
\end{abstract}

\maketitle

\section{Introduction}
Experimental investigations of fluctuations of conserved charges,
expressed as cumulants of net-particle multiplicity distributions,
probe the response of the system to external perturbations. For
example, the liquid gas phase transition can be probed by the response
of the volume to a change in pressure, which is encoded in the
isothermal compressibility.  Such measurements are hence particularly
interesting for studies of possible critical phenomena and the
existence of a critical endpoint in the QCD phase diagram~\cite{CP1, Misha, CP2}.  To make
any quantitative headway the objective is to isolate, in the
experimental data, the dynamical part of the fluctuations and compare
the corresponding cumulants to those from predictions for a thermal
system as obtained by calculations within the framework of lattice
gauge theory or other dynamical theories.

Indeed, for a thermal system of volume $V$ and temperature $T$, within
the Grand Canonical Ensemble, fluctuations of a given net-charge
$\Delta N_{B}=N_{B}-N_{\bar{B}}$ are related to the corresponding
reduced susceptibility $\hat{\chi}_{2}^{B}$~\cite{Bazavov, fluct_exp}:

\begin{equation}
\frac{1}{VT^{3}} \left(\left<\Delta N_{B}^{2}\right> - \left<\Delta N_{B}\right>^{2}\right)  = \hat{\chi}_{2}^{B},
\label{cumulants}
\end{equation}
with $\hat{\chi}_{2}^{B}$ defined as the second derivative of the
reduced thermodynamic pressure $\hat{p}\equiv \frac{p}{T^{4}}$ with
respect to the corresponding reduced chemical potential
$\hat{\mu}_{B}\equiv\frac{\mu_{B}}{T}$

\begin{equation}
 \hat{\chi}_{2}^{B} = \frac{\partial^{2}{\hat{p}}}{\partial \hat{\mu}_{B}^{2}}.
\label{pressure}
\end{equation}

In a similar way, higher order cumulants are related to the
corresponding higher order susceptibilities~\cite{Cheng}.  This means
that the response function of the system to external parameters can be
obtained from thermal averages of macroscopic variables by employing
the probability distribution of micro-states of the
system. Furthermore, in order to get rid of not directly measurable
quantities such as volume and temperature, which enter into
eq.~\ref{cumulants}, it is advocated in~\cite{Gupta} to look for
ratios of cumulants. However, a comment is in order here:
eq.~\ref{cumulants} is derived under the assumption that the volume of
the system is fixed. Within the Wounded Nucleon Model this means that
the number of participants is fixed in each event. In experiments,
however, events are classified into centrality bins. In most
theoretical approaches, centrality is specified using the collision
impact parameter; zero for central collisions, and close to the sum of
the radii of the colliding nuclei for peripheral collisions. More
specifically, one can define a centrality window incorporating the
$n\%$ most central collisions. Experimentally, however, one does not
have direct access to the impact parameter, hence the centrality
classes are typically defined as windows of energy deposited in a
zero-degree calorimeter, the number of participants, the multiplicity
of charged particles produced in a given acceptance, etc.

For the analysis of average quantities it is often not critical which
of the centrality determination approaches are used, because all of
them give similar results for such physical quantities.  However, the
situation changes dramatically if one considers event-by-event fluctuations
of these quantities. In this case, the centrality determination details
become crucial and differently influence the magnitude of measurements
of moments such as described in eq.~\ref{cumulants}. It is, therefore,
important to subtract from experimentally measured cumulants the
contributions originating from the fluctuations in the number of
wounded nucleons.

One should also note that fluctuation signals of conserved quantities
such as the net-baryon number need to be studied in a restricted phase
space~\cite{Koch_conserv}. Otherwise there are no fluctuations.  This
is achieved by placing appropriate cuts in rapidity and/or transverse
momentum of the detected particles. By construction, the smaller the
acceptance the smaller the effect of global conservation
laws. However, a too small acceptance may also destroy the
fluctuations of interest if the acceptance window is less than the
intrinsic dynamical correlation length $\lambda$~\cite{Misha}. This
issue becomes more and more important as one approaches a critical
endpoint in the QCD phase diagram, where the correlation length
becomes large~\footnote{For a system of infinite volume it diverges
there, see~\cite{Misha}.}. Clearly there, one should not select a
too small acceptance window. At low center-of-mass energies of
$\sqrt{s_{NN}} < 17$ GeV (corresponding to top SPS energy) the
(pseudo)rapidity width of the distribution of produced baryons $\Delta
\eta_B$ becomes smaller than the typical correlation length
$\lambda_{\eta} \approx 1$ (both expressed in terms of standard
deviations) even if there is no critical endpoint nearby, and effects
of baryon number conservation are expected to become dominant.

In this paper we address both of the effects mentioned above. By
simulating the actually used centrality selection criteria in the
ALICE experiment at the CERN LHC and the STAR experiment at the BNL
RHIC, we provide a framework to study the effects of participant
fluctuations on measured cumulants of any order and provide numerical
estimates on cumulants up to the order of four. Likewise, we give
simple but quantitative estimates of the corrections due to baryon
number conservation on measured cumulants.

We would like to mention that the contributions of participant or
'volume' fluctuations to dynamical event-by-event fluctuation signals
have been investigated previously in different
contexts~\cite{SI,Begun}. Moreover, the authors of~\cite{Skokov}
studied the effects of volume fluctuations on susceptibilities.
However, none of the previous studies have employed a detailed implementation of the experimentally used
centrality measures, a crucial ingredient of our approach.

Our paper is organized in the following way: first we collect and
summarize the notations and definitions used to compute cumulants. The
next two sections deal with participant or volume fluctuations and the
description of a simple model in which the effects of participant
fluctuations can be quantitatively simulated. We apply these
considerations in the following sections first to data from the ALICE
experiment at the LHC, followed by applications to selected STAR data
from the RHIC beam energy scan (BES). Next we discuss how to correct cumulant data for
the effect of global conservation laws. In the final section we
provide a conclusion and outlook.

\section {Notation and definitions}
In the following we choose the notations as used
e.g. in~\cite{definitions}.  The $r^{th}$ central moment of a discrete
random variable $X$, with its probability distribution $P(X)$, is
generally defined as

\begin{equation}
\mu_{r} \equiv  \left<\left(X-\left<X\right>\right)^{r}\right> = \sum_{X}\left(X-\left<X\right>\right)^{r}P(X),\\
\label{cum_definition1}
\end{equation}

where $\left<X\right>$ denotes the mean of the distribution

\begin{equation}
\left<X\right> =  \sum_{X}XP(X) .\\
\label{mean_definition1}
\end{equation}

In a similar way we introduce moments about the origin, thereafter referred to as raw moments 

\begin{equation}
 \left<X^{r}\right> = \sum_{X}X^{r}P(X).\\
\label{mean_definition2}
\end{equation}

Furthermore, to avoid particular units we introduce dimensionless moment ratios

\begin{equation}
\frac{\mu_{r}}{\left(\sqrt{\mu_{2}}\right)^{r}} =  \frac{\mu_{r}}{\sigma^{r}}, \\
\label{moments_ratios}
\end{equation}

where $\sigma^{2}$ is the variance defined by

\begin{equation}
\sigma^{2} \equiv \mu_{2}.  \\
\label{variance}
\end{equation}

For $r=3$,  eq.~(\ref{moments_ratios})  yields  the skewness of the distribution

\begin{equation}
\gamma_{1} \equiv   \frac{\mu_{3}}{\mu_{2}^{3/2}}  = \frac{\mu_{3}}{\sigma^{3}}.\\
\label{skew_definition}
\end{equation}
The quantity 'skewness' is a way to describe the asymmetry of a
particular distribution. The distribution is said to have positive
skewness if it has a longer tail to values larger (to the right) than
the central maximum compared to the left ones. If the reverse is true
the skewness is negative. On the other hand, the skewness is zero if
the data are symmetrically distributed about the mean.

The kurtosis of the distribution of $X$ is obtained by taking $r$ = 4
in eq.~(\ref{moments_ratios}),

\begin{equation}
\beta_{2}  \equiv   \frac{\mu_{4}}{\mu_{2}^{2}}  = \frac{\mu_{4}}{\sigma^{4}}.\\
\label{kurtosis_definition1}
\end{equation}
The 'kurtosis' is the degree of peakedness of the distribution,
usually taken relative to a normal distribution.  Since, for a normal
distribution, the value of the kurtosis is 3, it is usual to redefine
it as

\begin{equation}
\gamma_{2}  \equiv  \beta_{2} - 3 = \frac{\mu_{4} - 3\mu_{2}^{2}}{\sigma^{4}},\\
\label{kurtosis_definition2}
\end{equation}
which is generally referred to as kurtosis (sometimes it is called
kurtosis excess). Positive values of $\gamma_{2}$ imply a relatively
narrower peak and wider wings than the normal distribution with the
same mean and variance, while negative $\gamma_{2}$ values imply a
wider peak and narrower wings.

The cumulants of $X$ are defined as the coefficients in the Maclaurin
series of the logarithm of the characteristic function of $X$.  The
first four cumulants read
 \eq{\label{kumulants_definition}
  &\kappa_{1} = \left<X\right> , \nonumber\\ &\kappa_{2} = \mu_{2} =
  \left<X^{2}\right> - \left<X\right>^{2} , \nonumber\\ &\kappa_{3} =
  \mu_{3} = \left<X^{3}\right> - 3\left<X^{2}\right>\left<X\right> +
  2\left<X\right>^{3}, \\ &\kappa_{4} = \mu_{4} - 3\mu_{2}^{2} =
  \left<X^{4}\right> - 4\left<X^{3}\right>\left<X\right> -
  3\left<X^{2}\right>^{2} \nonumber\\ &
  +12\left<X^{2}\right>\left<X\right>^{2} -
  6\left<X\right>^{4}. \nonumber }
From eqs.~(\ref{skew_definition}), ~(\ref{kurtosis_definition2}) and
~(\ref{kumulants_definition}) we obtain the following relations which
are widely used in fluctuation analyses of conserved charges:

\begin{equation}
\gamma_{1}\sigma = \frac{\kappa_{3}}{\kappa_{2}},\\
\label{mean_definition3}
\end{equation}

and 

\begin{equation}
\gamma_{2}\sigma^{2} = \frac{\kappa_{4}}{\kappa_{2}}.
\label{mean_definition4}
\end{equation}
Finally, for the Poisson distribution, all its cumulants are equal to
its mean. The probability distribution of the difference $X_{1}$ -
$X_{2}$ of two random variables, each generated from statistically
independent Poisson distributions, is called the Skellam distribution.
According to the additivity of cumulants, the cumulants of the Skellam
distribution will then be

\begin{equation}
\kappa_{n}(Skellam)  = \left<X_{1}\right> + (-1)^{n}\left<X_{2}\right>,
\label{Skellam_definition}
\end{equation}
where $\left<X_{1}\right>$ and $\left<X_{2}\right>$ are mean values of  $X_{1}$ and $X_{2}$ respectively.

\section{Participant or Volume Fluctuations}
\label{volume_fluct}

Experimentally measured dynamical event-by-event fluctuation signals
such as cumulants of net-particle distributions can, as we will
demonstrate quantitatively below, be considerably modified by the
fluctuations of the target and projectile participants for a given
centrality selection.  In this section we will study the fluctuations
of participants within the framework of the Wounded Nucleon Model
(WNM)~\cite{WNMmodel}. We note that, in the WNM, particles are
produced from independent exited states of the nucleons (thereafter
referred to as sources, wounded nucleons, participants or
mini-fireballs). Each source can produce a number of particles,
however there are no correlations between different sources. Both
particles and antiparticles are produced from each source with the
same probability distribution, \emph{i.e.}, all sources are statistically
identical. We introduce the moment generating function for the net
particle $\Delta n = n_{B} - n_{\bar{B}}$ distributions from each
source $M_{\Delta n}(t)$, the exact expression of which is irrelevant
for our studies. Here $t$ is an auxiliary parameter, which is set to
zero after taking corresponding derivatives with respect to it.  The
raw moments of net-particles from each source can then be calculated
as

\begin{equation}
\left<\Delta n^{r}\right> = \left[\frac{d^{r}}{dt^{r}}M_{\Delta n}(t)\right]_{t=0}.
\label{WNM_1}
\end{equation}

On the other hand, the number of net-particles $\Delta N = N_{B} -
N_{\bar{B}}$ in a given event is a sum over net-particles from all
sources, within this event. As all sources are statistically
independent the moment generating function for the distribution of
$\Delta N$ will be equal to the product of the moment generating
functions from each source

\begin{equation}
M_{\Delta N}(t) = \left[M_{\Delta n}(t)\right]^{N_{W}},
\label{WNM_2}
\end{equation}
where $N_{W}$ is the number of sources which we take fixed for the
moment.

It is then straightforward to calculate any moments of the $\Delta N$
distribution. For example, for the first and the second raw moment we
obtain

\begin{equation}
\left<\Delta N\right>_{f} = \left[\frac{dM_{\Delta N}(t)}{dt}\right]_{t=0} = \left[N_{W}\left[M_{\Delta n}(t)\right]^{N_{W}-1}\frac{dM_{\Delta n}(t)}{dt}\right]_{t=0} = N_{W}\left<\Delta n\right>
\label{WNM_3}
\end{equation}

and

\begin{equation}
\left<\Delta N^{2}\right>_{f} = \left[\frac{d^{2}M_{\Delta N}(t)}{dt^{2}}\right]_{t=0}  = N_{W}\left(N_{W}-1\right)\left<\Delta n\right>^{2}+N_{W}\left<\Delta n^{2}\right>,
\label{WNM_4}
\end{equation}
where the index $f$  refers to the fixed number of $N_{W}$ and by definition $M_{\Delta n}(0) = 1$.

For a fluctuating number of wounded nucleons an additional summation
over the probability distribution of wounded nucleons $P(N_{W})$ is
needed

\begin{equation}
 \left<\Delta N\right>  = \sum_{N_{W}} \left<\Delta N\right>_{f} P(N_{W}) = \left<N_W\right>\left<\Delta n\right>
\end{equation}

and 

\begin{equation}
 \left<\Delta N^2\right>  = \sum_{N_{W}} \left<\Delta N^{2}\right>_{f} P(N_{W}) = \left<N_{W}\left(N_{W}-1\right)\right>\left<\Delta n\right>^{2}+\left<N_{W}\right>\left<\Delta n^{2}\right>.
\end{equation}

In a similar way any higher moments can be calculated. Finally,
substituting the so obtained raw moments into the above definitions of
cumulants (cf. eq.~\ref{kumulants_definition}) we get the following
expressions for the first four cumulants

\eq{
\label{cum1}
&\kappa_{1}(\Delta N) = \left<N_{W}\right>\kappa_{1}(\Delta n),\\
\label{cum2}
&\kappa_{2}(\Delta N) = \left<N_{W}\right>\kappa_{2}(\Delta n) + \left<\Delta n\right>^{2}\kappa_{2}(N_{W}),\\
\label{cum3}
&\kappa_{3}(\Delta N) =\left<N_{W}\right> \kappa_{3}(\Delta n) +   3\left<\Delta n\right>\kappa_{2}(\Delta n)\kappa_{2}(N_{W}) + \left<\Delta n\right>^{3}\kappa_{3}(N_{W}),\\
\label{cum4}
&\kappa_{4}(\Delta N) = \left<N_{W}\right>\kappa_{4}(\Delta n) +
4\left<\Delta n\right>\kappa_{3}(\Delta n)\kappa_{2}(N_{W}) \\ &+
3\kappa_{2}^{2}(\Delta n)\kappa_{2}(N_{W}) + 6\left<\Delta
n\right>^{2}\kappa_{2}(\Delta n)\kappa_{3}(N_{W}) + \left<\Delta
n\right>^{4}\kappa_{4}(N_{W}).\nonumber } 
Here, $\Delta n = n_{B} - n_{\bar{B}}$ is the number of net-particles
from a single source.  We note that the corresponding cumulants for
particles can be obtained by replacing $\Delta N$ and $\Delta n$ in
eqs.~\ref{cum1}-~\ref{cum4} with $N_{B}$ and $n_{B}$, respectively. In
the same way the cumulants for antiparticles are obtained.

As can be seen from the above equations, starting from the second
cumulants the fluctuations of the number of wounded nucleons which are
encoded in $\kappa_{n}(N_{W})$ modify the corresponding experimentally
measured cumulants $\kappa_{n}(\Delta N)$. Under the unrealistic
assumption of a fixed number of wounded nucleons ($\kappa_{n}(N_{W}) =
0, n > 1$) one obtains $\kappa_{n}(\Delta N) =
\left<N_{W}\right>\kappa_{n}(\Delta n)$, which implies that taking the
ratios of cumulants eliminates a dependence in the mean number of
wounded nucleons.  This particular case is then indeed equivalent to
eq.~\ref{cumulants}. This assumption is, however, not applicable for a
description of relativistic nuclear collisions, as mentioned above.

Interestingly, for LHC energies, fluctuations of $N_{W}$ are
irrelevant for $\kappa_{2}(\Delta N)$ and $\kappa_{3}(\Delta N)$.
This is because in eqs.~\ref{cum2} and ~\ref{cum3} the participant
fluctuation part scales with the mean number of net-particles $\left<\Delta
n\right>$ and its powers, which vanish for LHC energies at
mid-rapidity. This, however, does not hold for $k_{4}(\Delta N)$ and
all higher even cumulants. Indeed, taking $\left<\Delta n\right> = 0$
in Eq.~\ref{cum4} we get $k_{4}(\Delta N) =
\left<N_{W}\right>\kappa_{4}(\Delta n) + 3\kappa_{2}^{2}(\Delta
n)\kappa_{2}(N_{W})$. Nevertheless, for the fourth and higher even
cumulants the situation is much more favourable at LHC because some
contributions from higher cumulants of $N_{W}$ drop in these cases,
too.

\section{The Model}

In this section we develop a model to simulate the effects of participant or volume fluctuations 
 in the experimental environment and compare obtained results with the
equations derived in the previous section.  An important advantage of
this approach is that it allows a precise determination of the
statistics needed for a particular cumulant measurement, which is of
crucial importance for the preparation of an event-by-event
experiment. Further input from experimental data is necessary for a
successful analysis: (i) a detailed description of the centrality
selection procedure employed in a particular experiment, and (ii)
measurements of the first moments (mean multiplicities) of particles
and antiparticles.

As the centrality determination is a delicate experimental issue
(cf. the discussion in the introduction), each experiment has to be
considered separately. Below we implement one of the centrality
selection approaches used in the ALICE experiment, where the measured
multiplicities (signal amplitudes in VZEROs) are fitted with those
obtained from a Glauber Monte Carlo simulation (for details see~\cite{ALICE_CENTRALITY}).

\begin{figure}[htb]
 \includegraphics[width=1.0\linewidth,clip=true]{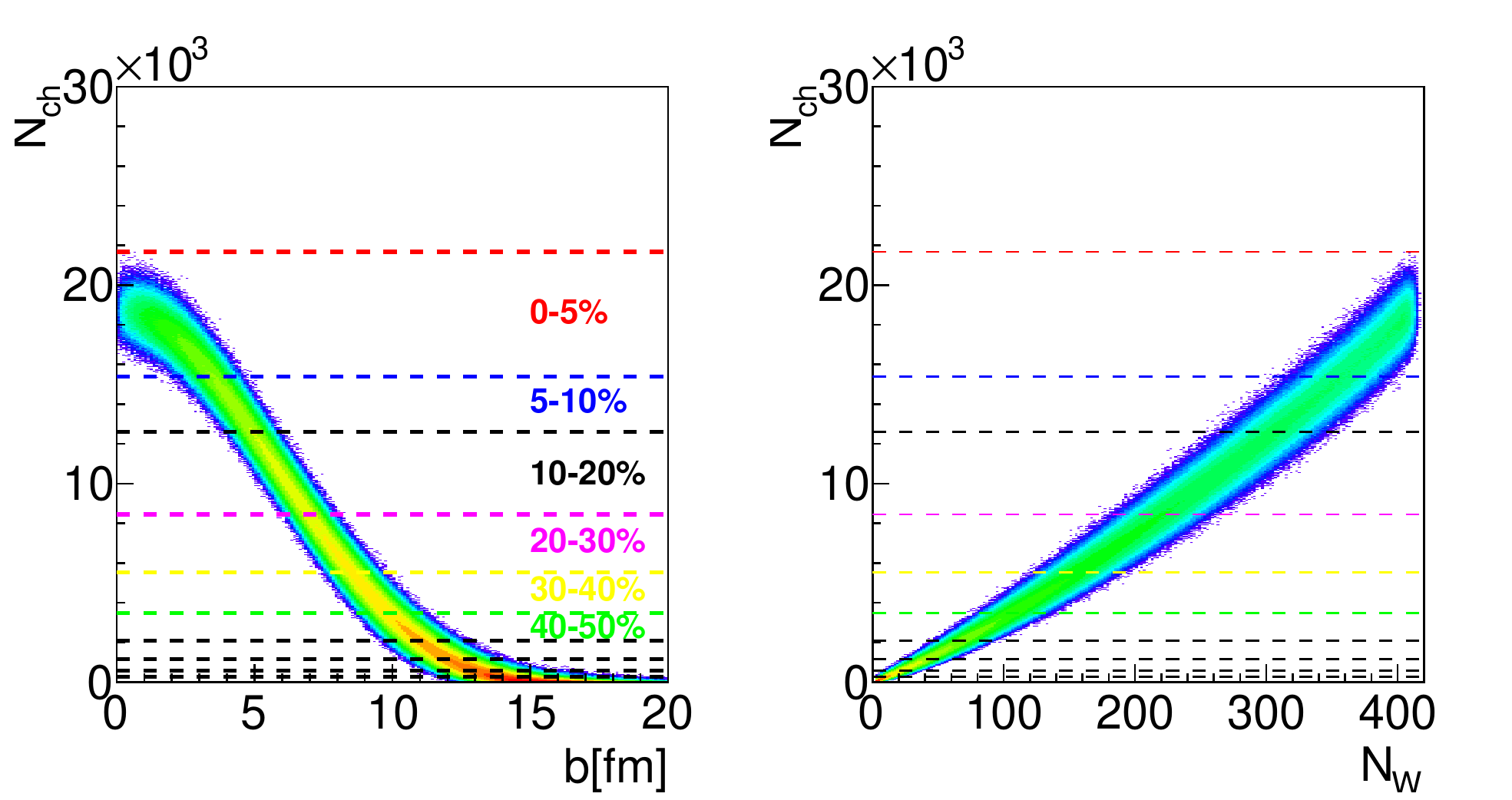}
 \caption{Left Panel: Produced number of charged particles versus the
   impact parameter. Right Panel: Produced number of charged particles
   versus the number of wounded nucleons. For a given value of the
   impact parameter the number of wounded nucleons and binary
   collisions are calculated with a Glauber Monte Carlo simulation
   based on the approach described in \cite{ALICE_CENTRALITY}. Next,
   using a two-component model, charged particles are produced
   assuming a Negative Binomial Distribution with parameters extracted
   by the same procedure as used in the ALICE experiment. }
\label{cent1}
\end{figure}   

Technically, following a two-component model~\cite{twoc1, twoc2}, in
which one decomposes nucleus-nucleus collisions into soft and hard
interactions, we first calculate the number of ancestors

\begin{equation}
N_{ancestors}=fN_{W}+(1-f)N_{coll},
\label{twocomponent}
\end{equation}
where $N_{W}$ and $N_{coll}$ are the number of wounded nucleons and
binary collisions, simulated in each Glauber Monte Carlo event for a
given value of the impact parameter~\cite{Glauber} and $f = 0.801$ is
taken from~\cite{ALICE_CENTRALITY}.

Next, from each ancestor we generate particles from a Negative
Binomial Distribution (NBD), defined by the probability distribution

\begin{equation}
P_{\mu,k}(n)=\frac{\Gamma(n+k)}{\Gamma(n+1)\Gamma(k)}\left(\frac{\mu}{k}\right)^{n}\left(\frac{\mu}{k} + 1\right)^{-(n+k)},
\label{NBD}
\end{equation}
where $\mu$ is the mean multiplicity of particles emitted from each
ancestor and $k$ controls the width of the NBD.  Numerical values of
the parameters, $\mu = 29.3$ and $k = 1.6$, are taken from the ALICE
paper~\cite{ALICE_CENTRALITY}.

\begin{figure}[htb]
 \includegraphics[width=1.0\linewidth,clip=true]{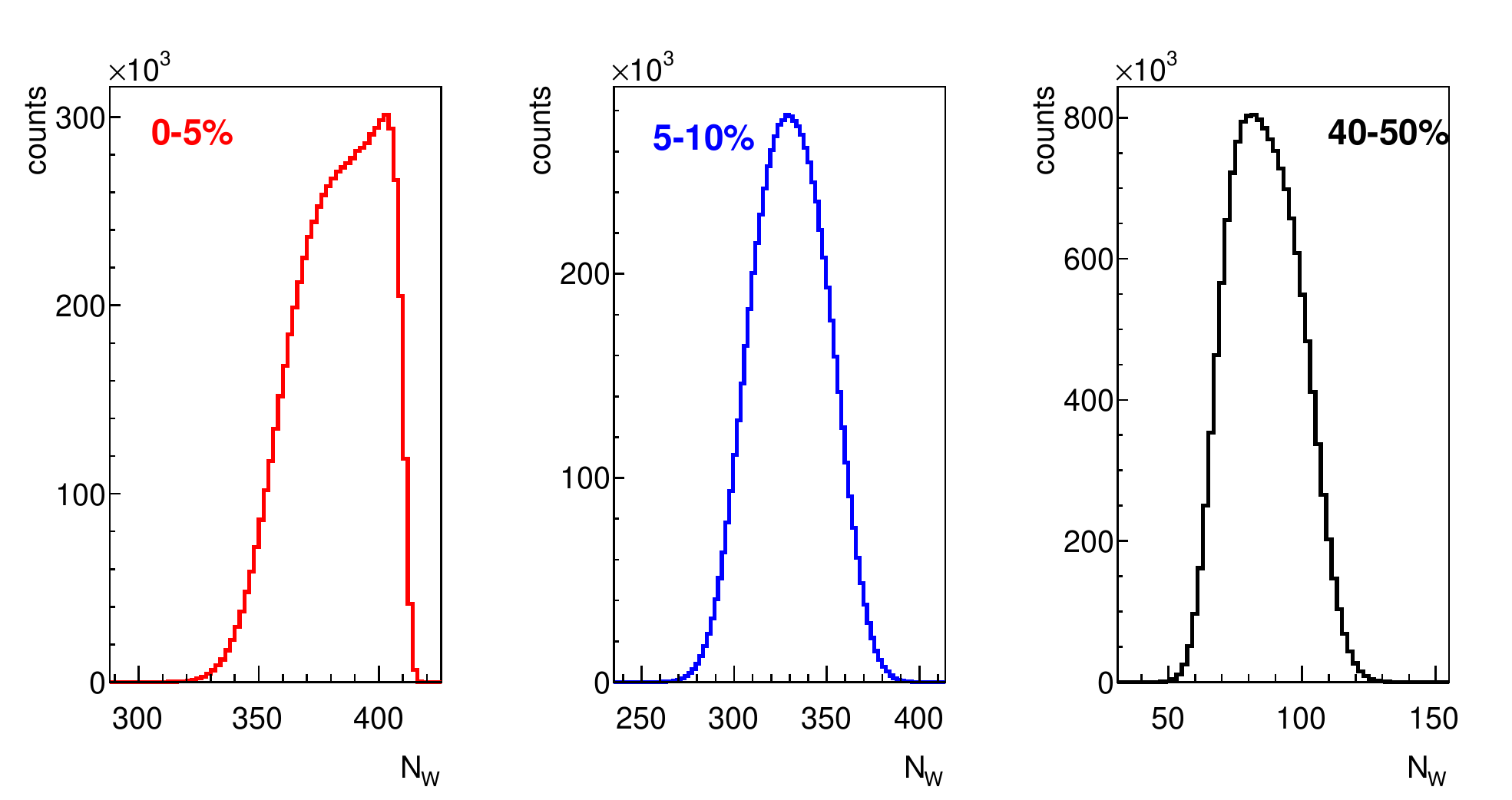}
 \caption{Distributions of the number of wounded nucleons for 3
   different centrality classes selected by applying sharp cuts on the
   number of produced charged particles. The left panel corresponds to
   selections of 0-5\%, the middle panel to 5-10\% and the right panel
   to 40-50\% of the total inelastic cross section.}
\label{cent2}
\end{figure}

Two-dimensional scatter plots representing the dependence on $b$ and
$N_{W}$ of the produced number of charged particles are presented in
the left and the right panel of Fig.~\ref{cent1}, respectively. The
centrality classes, selected by applying sharp cuts on the number of
produced charged particles (y axis), are represented by the dashed
horizontal lines. As seen from the scatter plots in the
Fig.~\ref{cent1}, where each dot represents one single event, the
impact parameter as well as the number of wounded nucleons fluctuate
from event-to-event, thus generating a distribution. To demonstrate
this explicitly we present, in Fig.~\ref{cent2}, distributions of
wounded nucleons for 3 different centrality classes.

For the $5\%$ most central collisions we observe that the
distribution is asymmetric and has a tail towards lower values of
wounded nucleons. This is caused by the fact that the number of
wounded nucleons cannot exceed two times the mass number of the
colliding nuclei, {\it i.e.} $416$, in the case of $Pb+Pb$
collisions. As a consequence, higher cumulants of the distribution of
wounded nucleons acquire large values for this centrality bin.  This,
in turn, distorts the experimentally measured cumulants of both
particles and net-particles. Indeed, according to
eqs.~\ref{cum2}-~\ref{cum4}, the cumulants of the participant
distributions $k_{n}(N_{W})$ are entering into the measured cumulants
of net-particles $k_{n}(\Delta N)$. In the following, we will study
these contributions based on experimental measurements of the
distributions of protons and antiprotons at the LHC and RHIC,
respectively.

\begin{figure}[htb]
 \includegraphics[width=1.0\linewidth,clip=true]{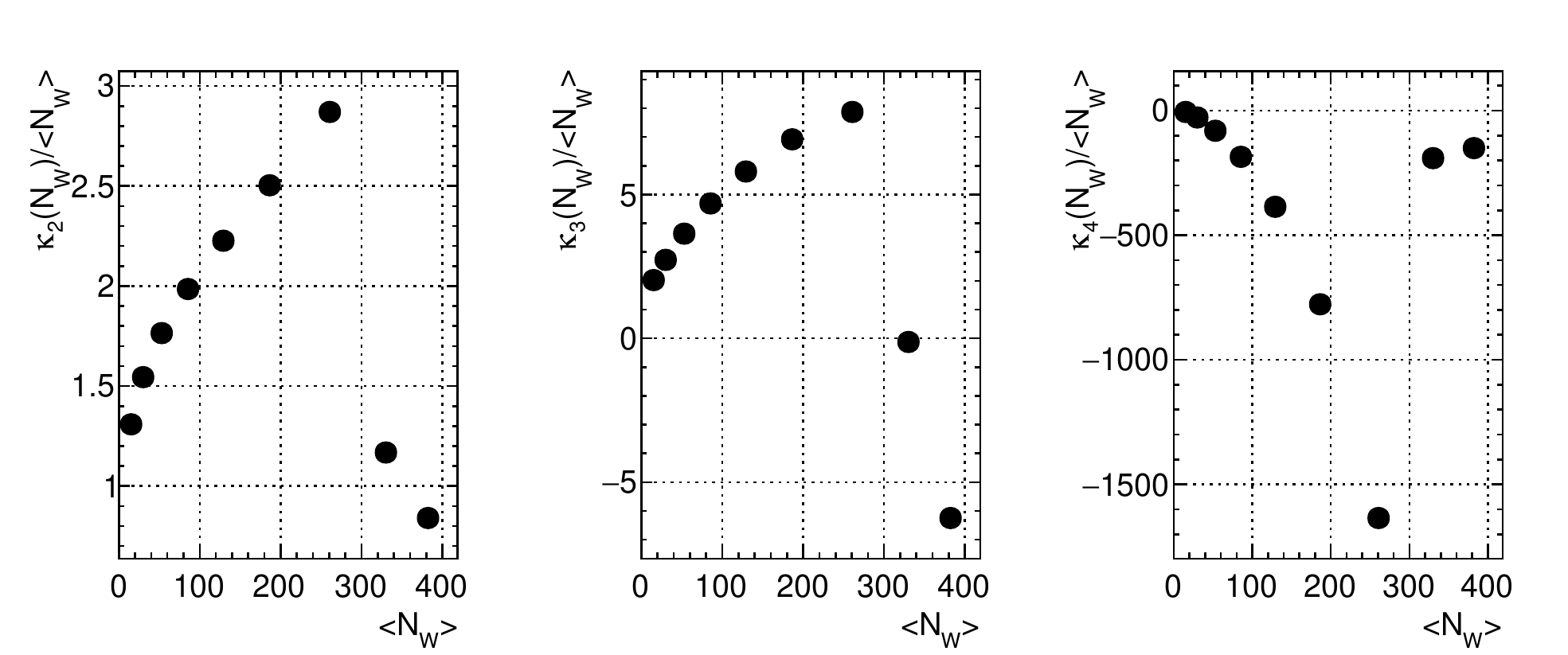}
 \caption{Normalised cumulants of wounded nucleon distributions as
   functions of the mean number of wounded nucleons.} 
\label{norm_cum}
\end{figure}   

\subsection{LHC energies}

The centrality dependence of cumulants of the wounded nucleon
distributions, normalised to the mean number of wounded nucleons, are
presented in Fig.~\ref{norm_cum}.

\begin{figure}[htb]
 \includegraphics[width=0.45\linewidth,clip=true]{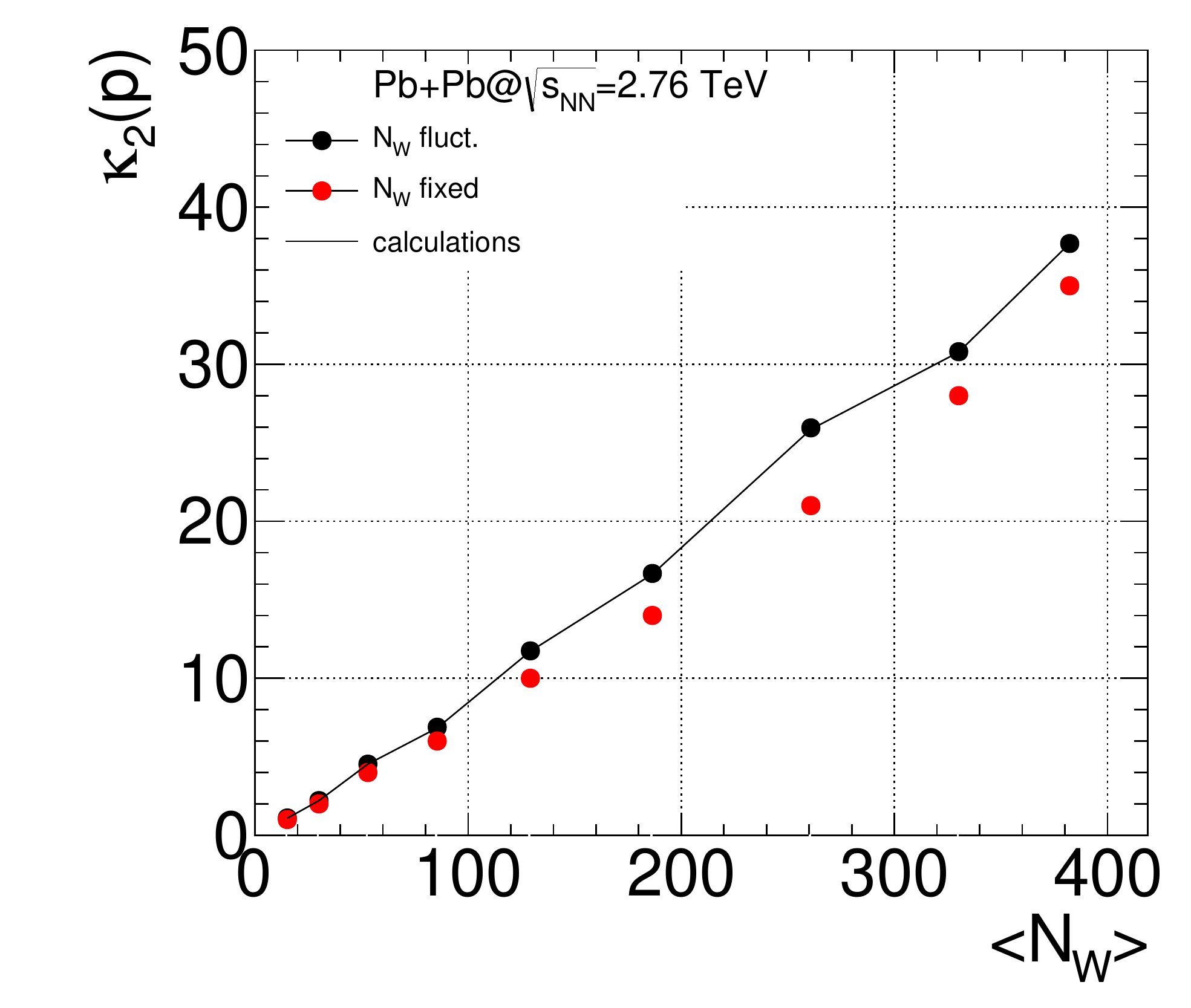}
 \includegraphics[width=0.45\linewidth,clip=true]{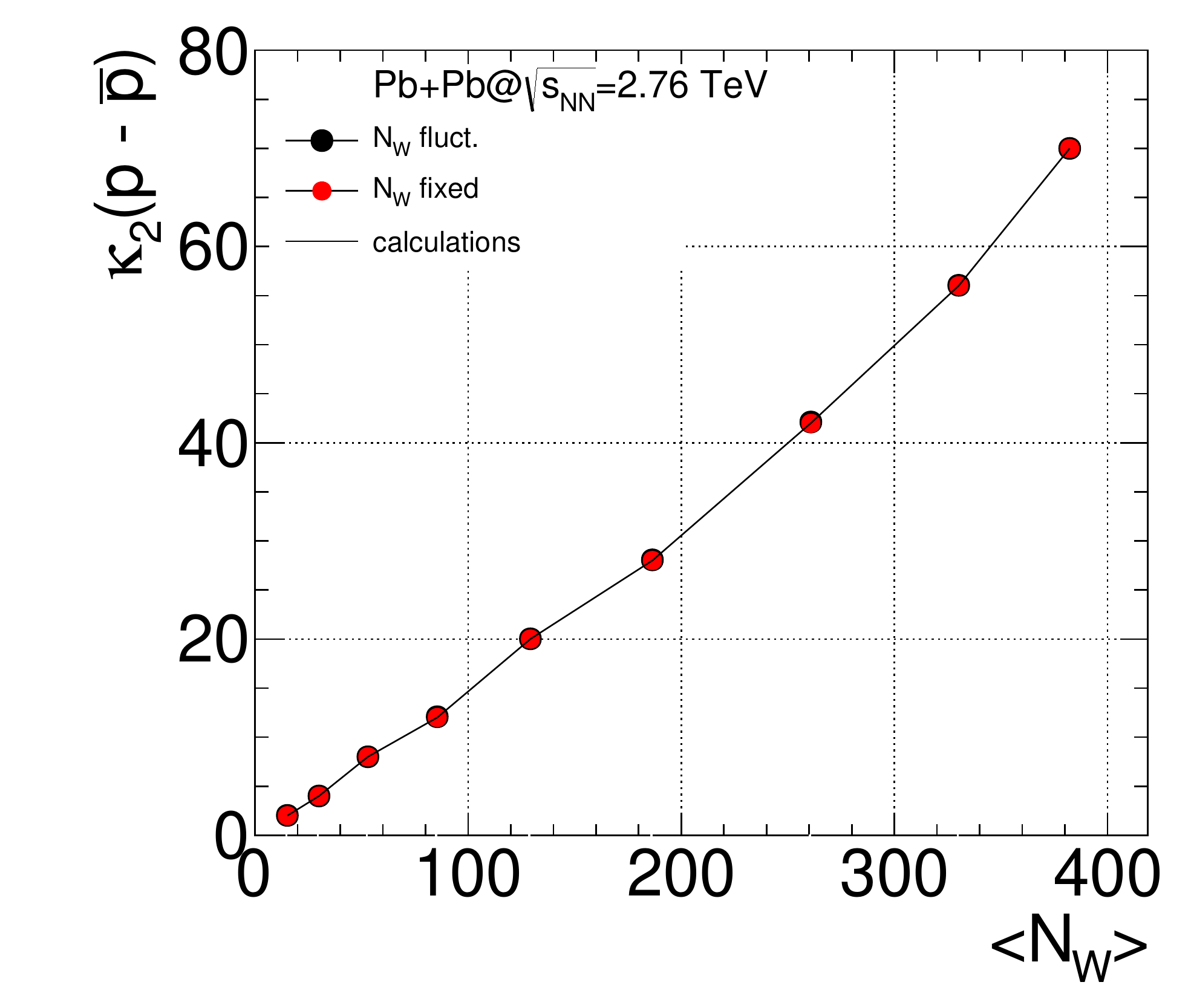}
 \caption{Second cumulants of protons (left panel) and net-protons
   (right panel). The red points correspond to a fixed number of
   wounded nucleons, while for the black points the fluctuations of
   wounded nucleons are included.  In the case of vanishing mean
   number of net-protons, their second cumulants do not depend on
   participant fluctuations (see text). The black lines are calculated using eq.~\ref{cum2}.}
\label{k2}
\end{figure}

\begin{figure}[htb]
 \includegraphics[width=0.45\linewidth,clip=true]{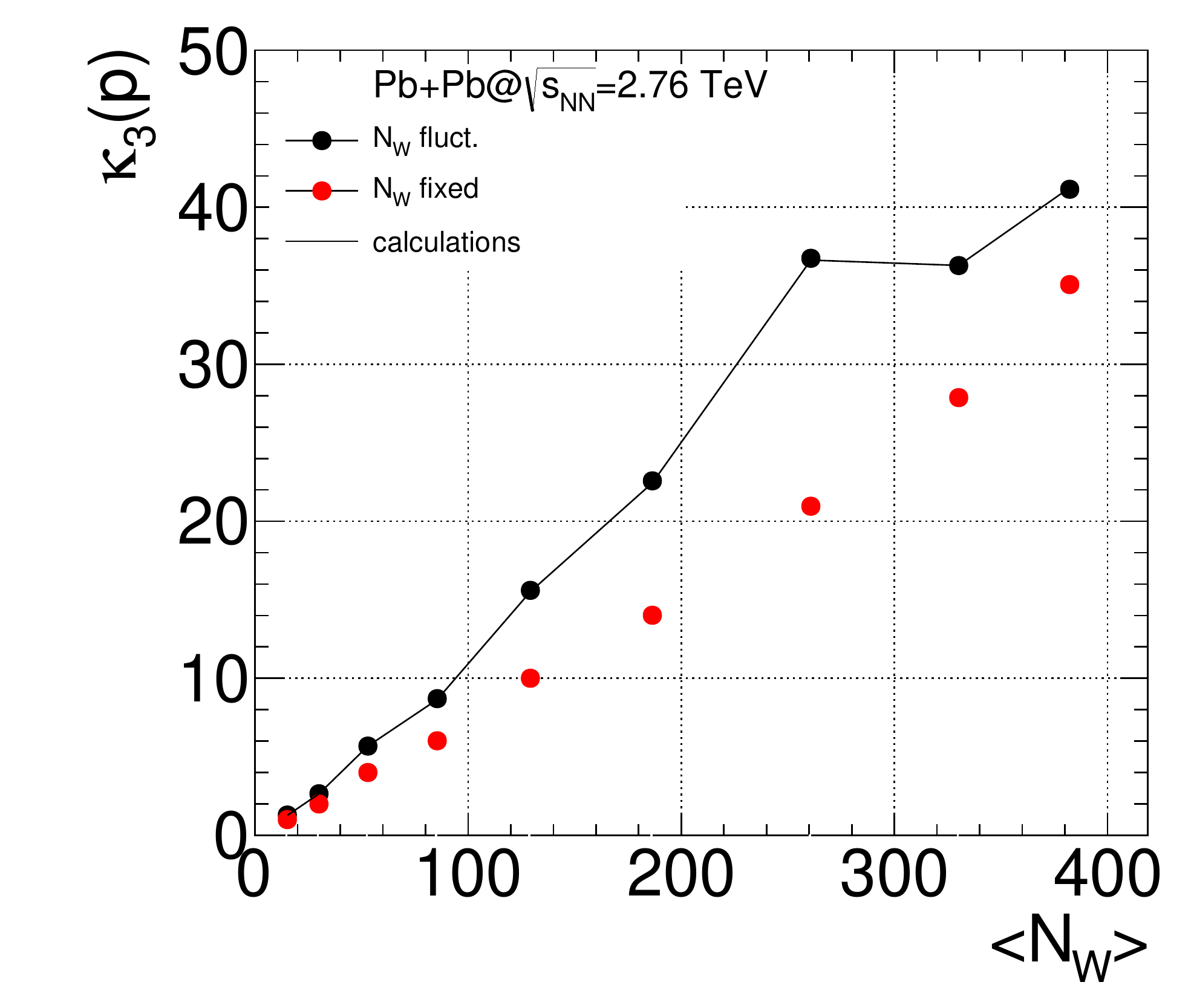}
 \includegraphics[width=0.45\linewidth,clip=true]{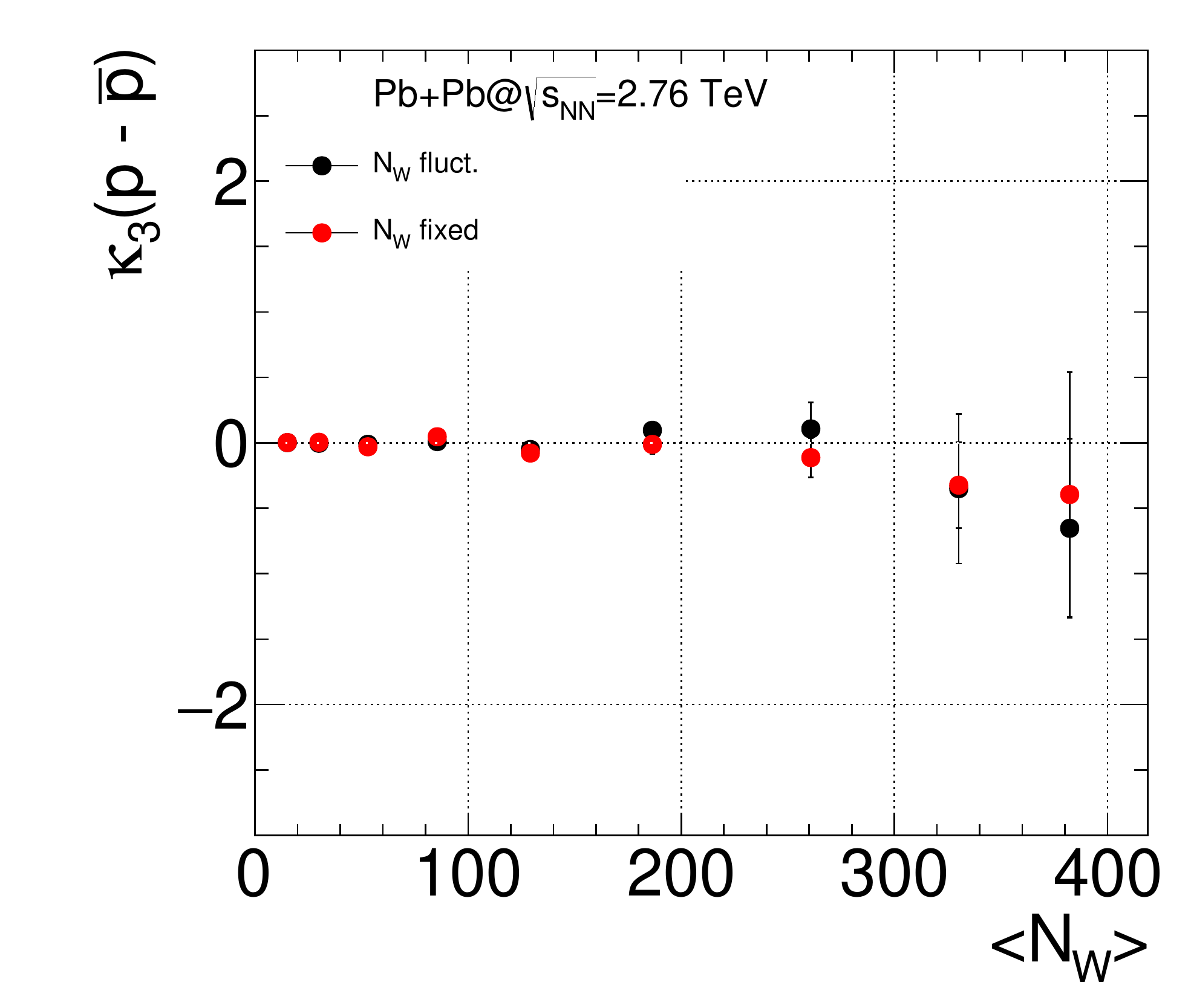}
 \caption{Third cumulants of protons (left panel) and net-protons
   (right panel). The red points correspond to a fixed number of
   wounded nucleons while, in the black points, fluctuations of
   wounded nucleons are included.  In the case of vanishing mean
   number of net-protons, their third cumulants do not depend on
   participant fluctuations. The black line is calculated using eq.~\ref{cum3}.}
\label{k3}
\end{figure}

\begin{figure}[htb]
 \includegraphics[width=0.45\linewidth,clip=true]{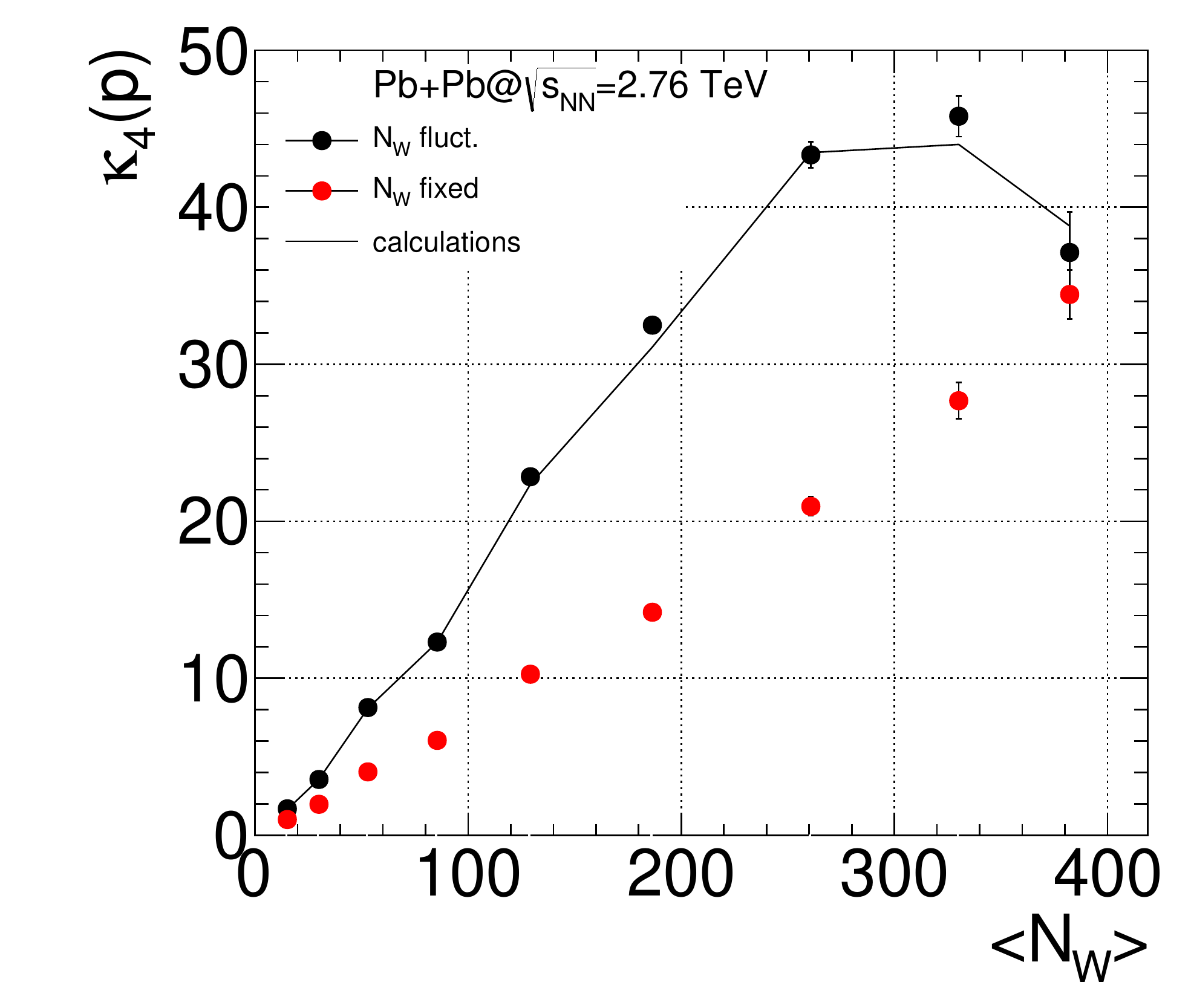}
 \includegraphics[width=0.45\linewidth,clip=true]{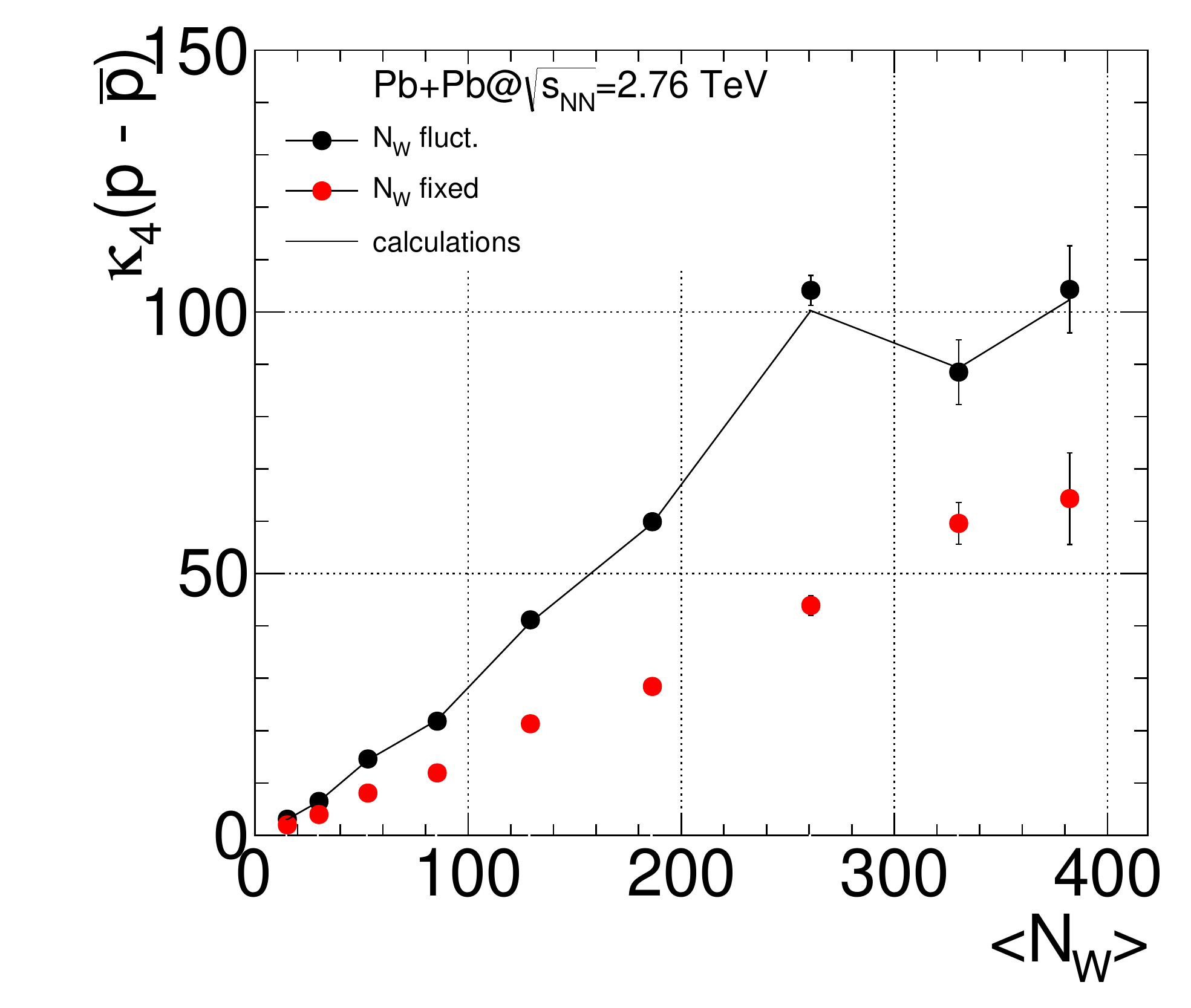}
 \caption{Fourth cumulants of protons (left panel) and net-protons
   (right panel). Red points correspond to the fixed number of wounded
   nucleons, while in black distributions fluctuations of wounded
   nucleons are included.  The black lines are calculated using eq.~\ref{cum4}. }
\label{k4}
\end{figure}

Protons and antiprotons are produced from each wounded nucleon or
'mini-fireball' within the framework of the Grand Canonical
Ensemble. In doing so we first define two independent Poisson
distributions; one for protons and another one for antiprotons.  The
mean values of protons and antiprotons, which define the corresponding
Poisson distributions, are taken from the ALICE measurements in
$Pb+Pb$ collisions at $\sqrt{s_{NN}}=2.76$
TeV~\cite{ALICE_Multiplicities}.  We assume that the first moment of
the net-proton distribution vanishes, hence we take the same mean
number of antiprotons and protons, which is quantitatively supported
by both experimental measurements~\cite{ALICE_Multiplicities} and
Hadron Resonance Gas model analysis, see \cite{HRG}. Next, for each
source we generate protons and antiprotons from independent Poisson
distributions. Each generated event is thus characterised by the
number of wounded nucleons $N_{W}$ (different for each event), as well
as the resulting number of protons and antiprotons. The event averages
of these quantities, with simulated $150\times 10^6$ events, expressed
in terms of cumulants of protons and net-protons are presented in
Figs.~\ref{k2}-~\ref{k4}. Here, red symbols represent results computed
under the assumption of keeping the number of wounded nucleons fixed,
while black symbols correspond to the full simulation, {\it i.e.},
wounded nucleons fluctuate from event to event. We note that as a
fixed number of wounded nucleons we used their values averaged over
all events. Black lines are calculated using eqs.~\ref{cum2}
-~\ref{cum4}, where for protons $\Delta N = N_{p}$ and $\Delta n =
n_{p}$ were used.
 
As is seen from the left panel of Fig.~\ref{k2}, the second cumulants
of protons are modified by the fluctuations of wounded
nucleons. Moreover, starting from the third centrality class, the
centrality bin width used in the data analysis is doubled (see
Fig.~\ref{cent1}), leading to enhanced fluctuations in the
distributions of wounded nucleons. This is the reason for the
kink-like structure in the centrality dependence of the second
cumulants for protons presented in the left panel of the
Fig.~\ref{k2}. The contribution from wounded nucleon fluctuations
to the second cumulants of protons, encoded in the second cumulants of
wounded nucleons $k_{2}(N_{W})$, are scaled with the square of the
mean number number of protons from each wounded nucleon
(cf. eq.~\ref{cum2}). On the other hand, as mentioned above, at ALICE
energies the mean number of protons and anti-protons are nearly
identical, see \cite{ALICE_Multiplicities} and \cite{HRG},
 implying that the participant fluctuation part for the second
cumulants of net-proton distributions vanishes. As a consequence, the
second cumulants of net-protons at ALICE energies are not affected by
the fluctuations of wounded nucleons, as demonstrated in the right
panel of the Fig.~\ref{k2}, where the red and black symbols coincide.

In Fig.~\ref{k3}, the centrality dependence for the third cumulants of
protons and net-protons are presented. We again observe strong
contributions from the fluctuations of wounded nucleons which are
evidenced by differences between the red and black distributions for
protons. Furthermore, the variation in the width of the centrality
class leads to an even more pronounced kink structure, which in turn
makes the centrality dependence non-monotonic. Fortunately, for the
third moments of the net-proton distributions at ALICE energies, the
contributions from participant fluctuations still vanish.
\begin{figure}[htb]
 \includegraphics[width=0.5\linewidth,clip=true]{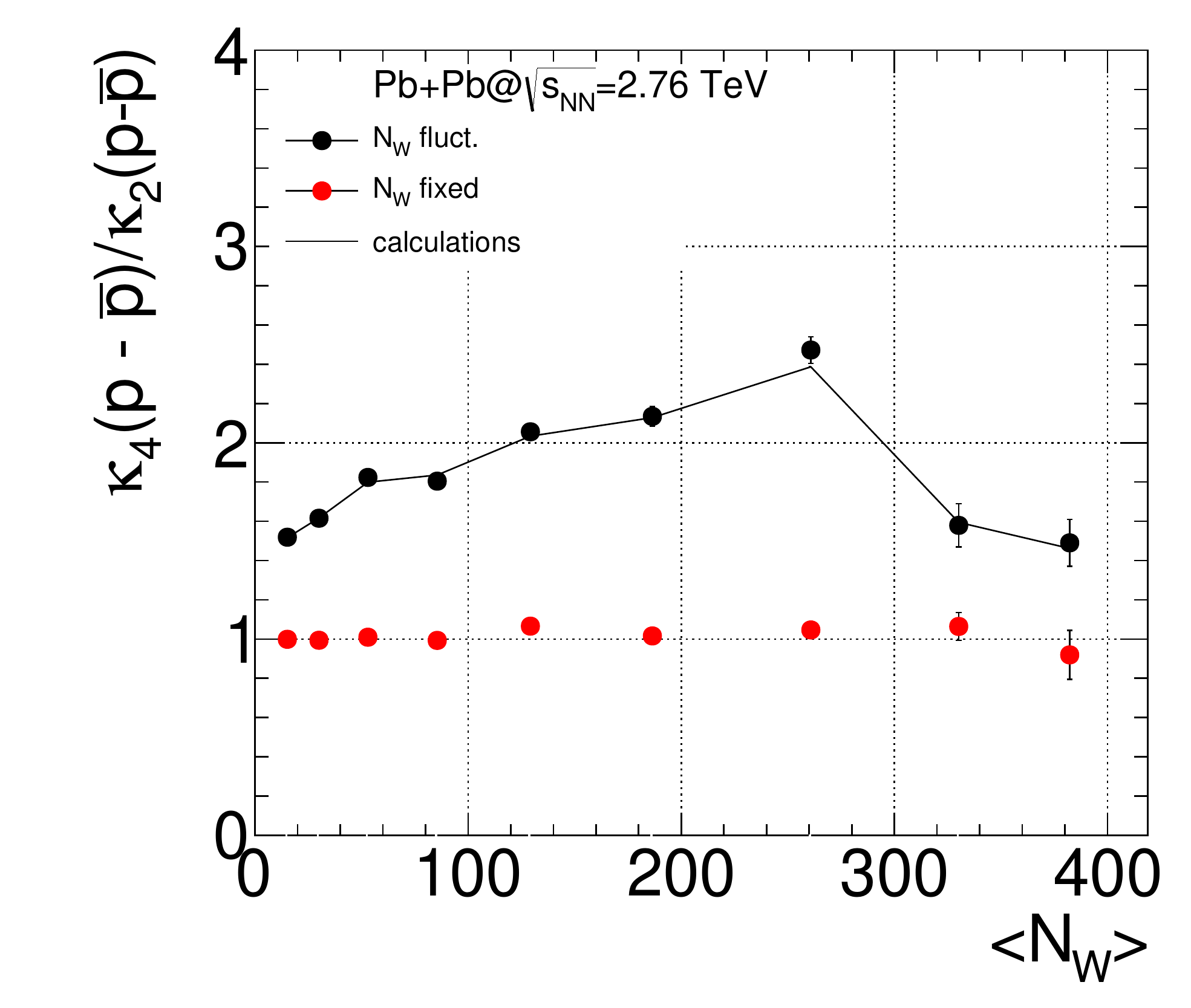}
 \caption{Ratio of cumulants for net-protons. Red points correspond to
   keeping the number of wounded nucleons fixed, while for the black
   points the fluctuations of wounded nucleons are included. The black line is calculated using eq.~\ref{cum4_net0_ratio}.}
\label{kRatio}
\end{figure}  
This is because, in eq.~\ref{cum3}, the cumulants of the participants
are scaled with the mean number of net-protons. The situation changes
significantly for the fourth cumulants of net-protons.  Our
calculations demonstrate that, for the fourth cumulants, the
contributions from participant fluctuations are not removed even for
vanishing mean value of net-particles. Indeed, by setting
$\left<\Delta n \right> = 0$ in eq.~\ref{cum4} we obtain for the
fourth cumulants of net-protons:
\begin{equation}
\kappa_{4}(\Delta N) = \left<N_{W}\right>\kappa_{4}(\Delta n) +   3\kappa_{2}^{2}(\Delta n)\kappa_{2}(N_{W}).
\label{cum4_net0}
\end{equation}
Here, the first term corresponds to the dynamical fluctuations we are
interested in. Unless corrected, this gets masked by the second term
that includes the second cumulant of participant distributions. The
latter is quite significant as seen in the right panel of
Fig.~\ref{k4}, where the black line (very close to the symbols) is
plotted using eq.~\ref{cum4_net0}. In the left panel of Fig.~\ref{k4}
fourth cumulants of protons are presented.  For the first centrality
bin (0-5$\%$) we observe rather small difference between red and black
symbols. This is because fourth cumulants of protons gets modified by
$\kappa_{2}(N_{W})$, $\kappa_{3}(N_{W})$ and $\kappa_{4}(N_{W})$
(cf. eq.~\ref{cum4}). On the other hand,
as seen in Fig.~\ref{norm_cum}, both $\kappa_{3}(N_{W})$ and
$\kappa_{4}(N_{W})$ are negative for this centrality bin. The
interplay between mean number of protons and cumulants of wounded
nucleons may by chance cancel the effect of volume fluctuations.  In
Fig.~\ref{kRatio} we present the ratio of the fourth and second
cumulants of net-protons. Since our simulations involve no dynamical
net-proton fluctuations, for fixed number of wounded nucleons the
ratio is unity (red points). However, due to participant fluctuations,
the results get modified by a factor of more than two (black points).
This can also be explained analytically by taking the ratio of the
corresponding cumulants in eqs.~\ref{cum4} and ~\ref{cum2} for
$\left<\Delta n \right> = 0$

\begin{equation}
\frac{\kappa_{4}(\Delta N)}{{\kappa_{2}(\Delta N)}} =  \frac{\kappa_{4}(\Delta n)}{{\kappa_{2}(\Delta n)}}  +   3\kappa_{2}(\Delta n)\frac{\kappa_{2}(N_{W})}{\left<N_{W}\right>}.
\label{cum4_net0_ratio}
\end{equation}
We thus obtain that, even at ALICE energies, the ratio of the fourth
to the second cumulants of net-protons is significantly modified by
fluctuations of participants scaled with the second cumulant of
net-protons. This implies that the enhancement of the fourth cumulant
of net-protons due to participant fluctuations will introduce a
significant bias into this cumulant ratio.

\subsection{RHIC energies}

We demonstrated in the previous section our results at  LHC energies, where
the mean number of net-protons is, to a high degree of accuracy, zero.
There the influence of participant fluctuations on the second and third net-
proton cumulants vanishes. The fourth (and all higher even )
cumulants, however, receive significant contributions from such fluctuations. At
lower energies the mean number of net-protons increases, and no such
cancellation is expected then even for the 2nd and 3rd cumulants.

\begin{figure}[htb]
 \includegraphics[width=0.45\linewidth,clip=true]{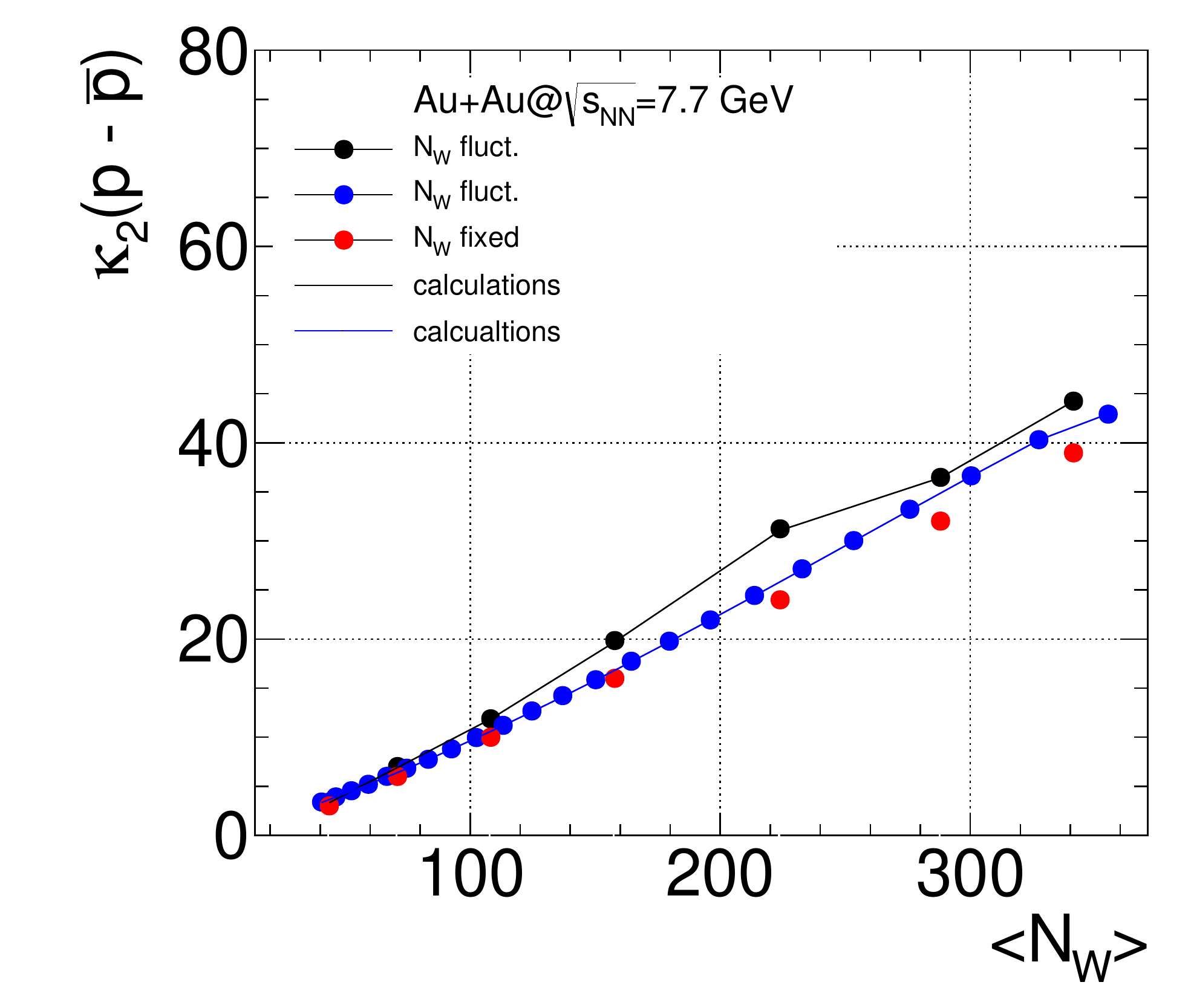}
 \includegraphics[width=0.45\linewidth,clip=true]{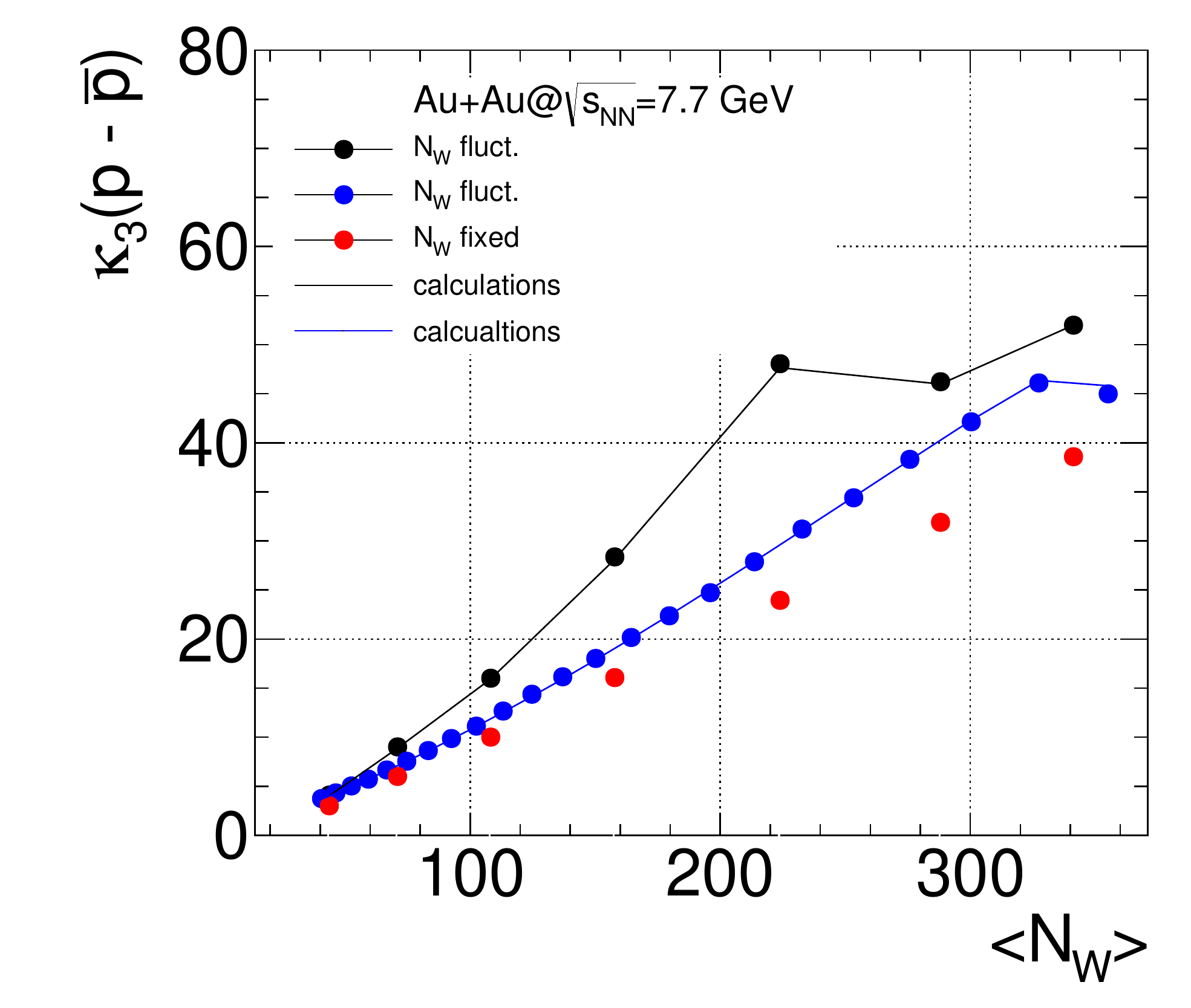}
 \caption{Second (left panel) and third (right panel) cumulants of
   net-protons for Au+Au collisions at $\sqrt{s_{NN}}$=7.7GeV. Red
   points correspond to keeping the number of wounded nucleons fixed,
   while for the blue and black points the fluctuations of wounded
   nucleons are included. The centrality bin width is $2.5\%$ for the
   blue points, while for the black points variable bin widths (see
   Fig.~\ref{cent1}) are used. The  lines (black and blue) are calculated using eqs.~\ref{cum2} and~\ref{cum3}.}
\label{star23}
\end{figure}  

To study this in detail we present, in this section, the results for
second, third and fourth cumulants of net-proton distributions for
Au-Au collisions at two different values of $\sqrt{s_{NN}}$, namely 39
and 7.7 GeV. The latter value is the lowest energy at which the STAR
collaboration has taken data in the framework of the RHIC
BES. To this end we have simulated $150\times 10^{6}$ minimum
bias Au+Au collisions at $\sqrt{s_{NN}}$ = 7.7 and 39 GeV.

In doing so we have neglected any possible correlations between
charged particles, employed for the centrality determination, and
those used for the event-by-event analysis. Such (auto-)correlations
are unavoidable if the rapidity window used for the centrality
determination is not sufficiently different from that used for the
fluctuation measurements. Mean multiplicities of protons are taken
from~\cite{STARDATA}, whereas for anti-protons they are set to zero.
As explained in~\cite{STARDATA}, the STAR experimental data points
have been modified by the so called Centrality Bin Width Correction
(CBWC)~\cite{CBWC}. The essential idea behind the CBWC is to get rid
of the participant fluctuations by subdividing a given centrality bin
into smaller ones and then merging them together incoherently. In
Fig.~\ref{star23} we present second and third cumulants of net-protons
as function of centrality, with variable (black points) and fixed bin
width of $2.5\%$ (blue points).

\begin{figure}[htb]
 \includegraphics[width=0.45\linewidth,clip=true]{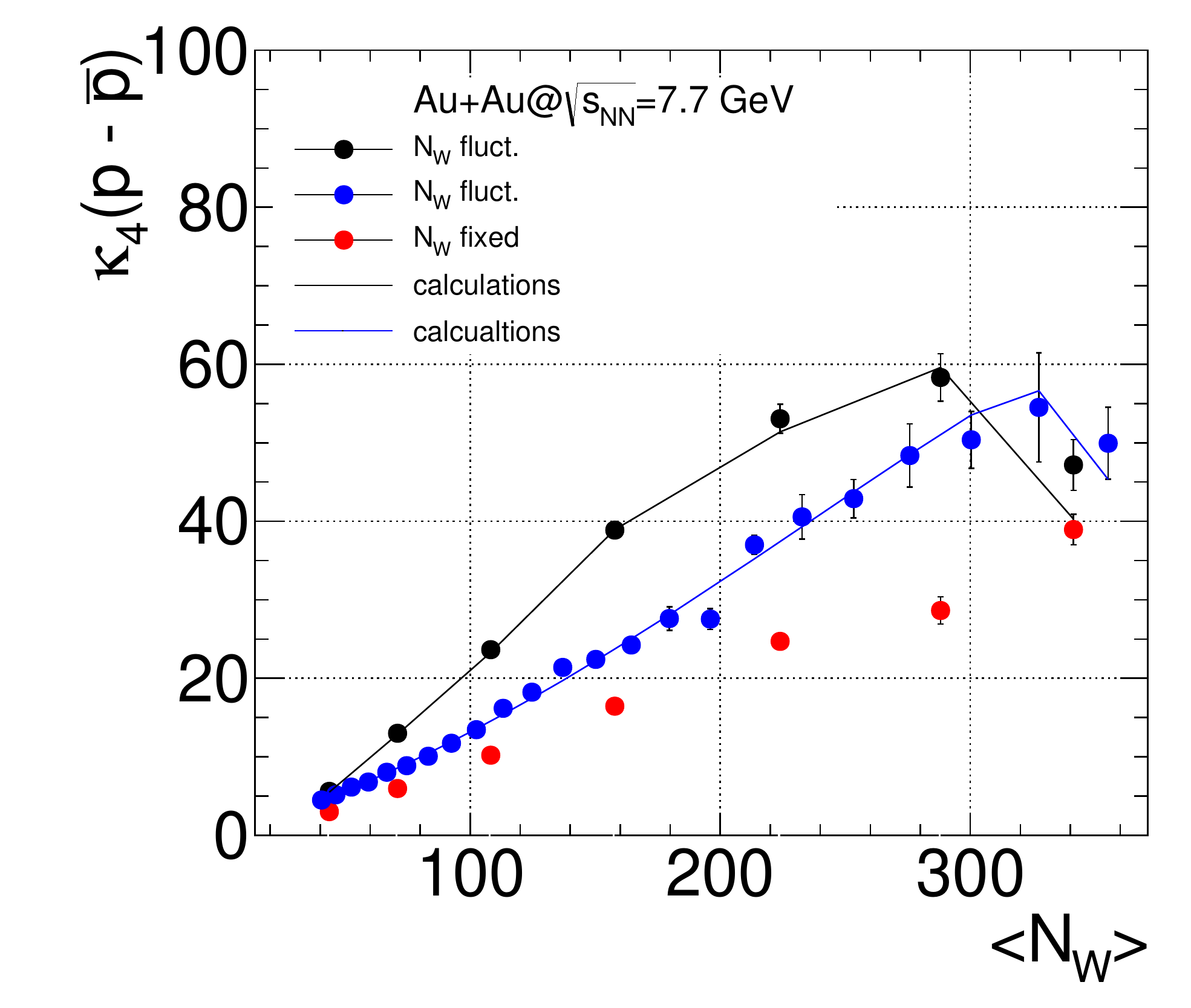}
 \includegraphics[width=0.45\linewidth,clip=true]{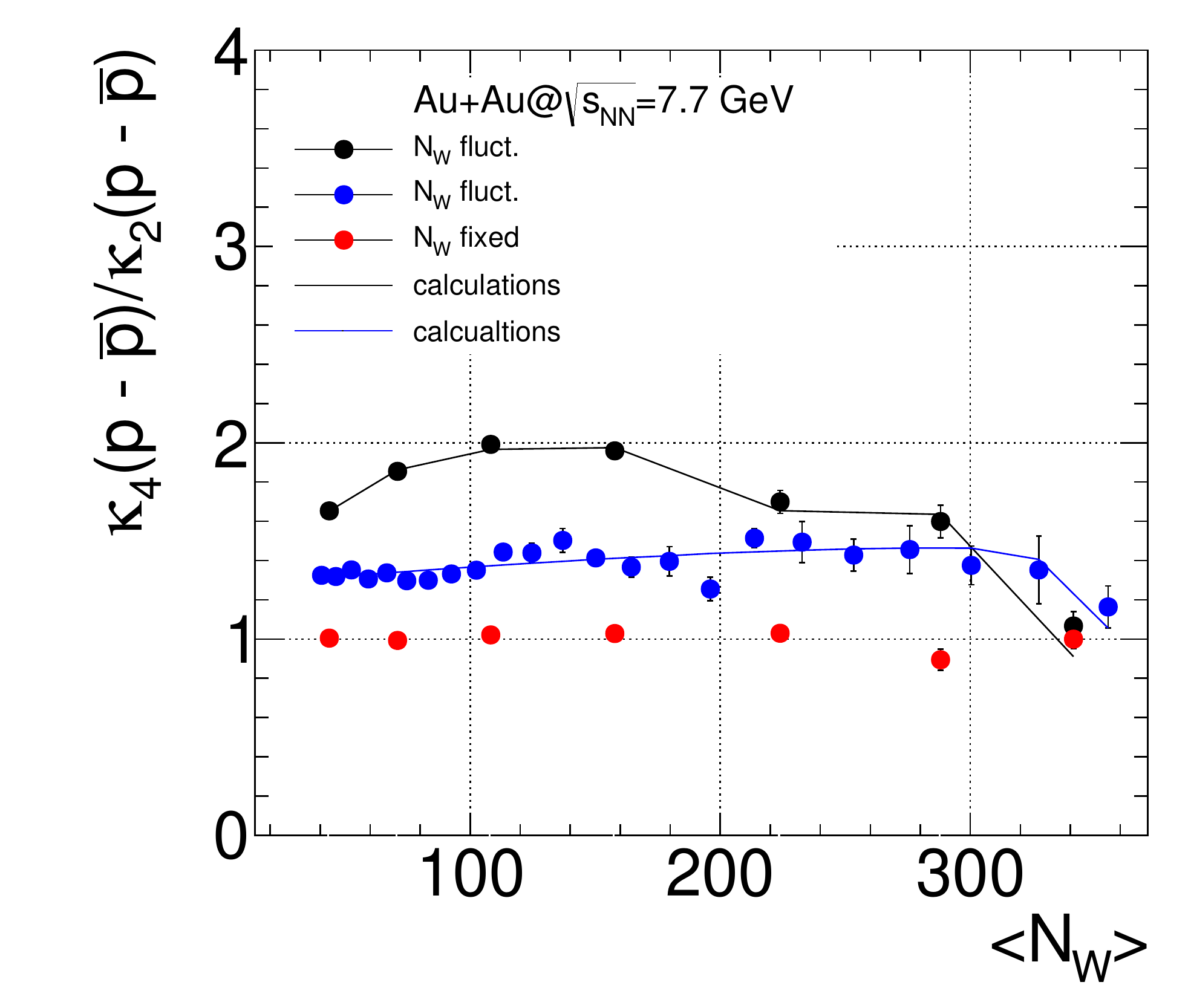}
 \caption{Left panel: Fourth cumulants of net-protons for Au+Au
   Collisions at $\sqrt{s_{NN}}=7.7GeV$. Right panel: Ratio of fourth
   and second cumulants.  Red points correspond to fixed number of
   wounded nucleons while, for the black points, the fluctuations of
   wounded nucleons are included. The centrality bin width is $2.5\%$
   for the blue points, while for the black points variable bin widths
   (see Fig.~\ref{cent1}) are used. The  lines (black and blue) are calculated using eqs.~\ref{cum2} and~\ref{cum4}.}
\label{star4}
\end{figure}  

\begin{figure}[htb]
 \includegraphics[width=0.45\linewidth,clip=true]{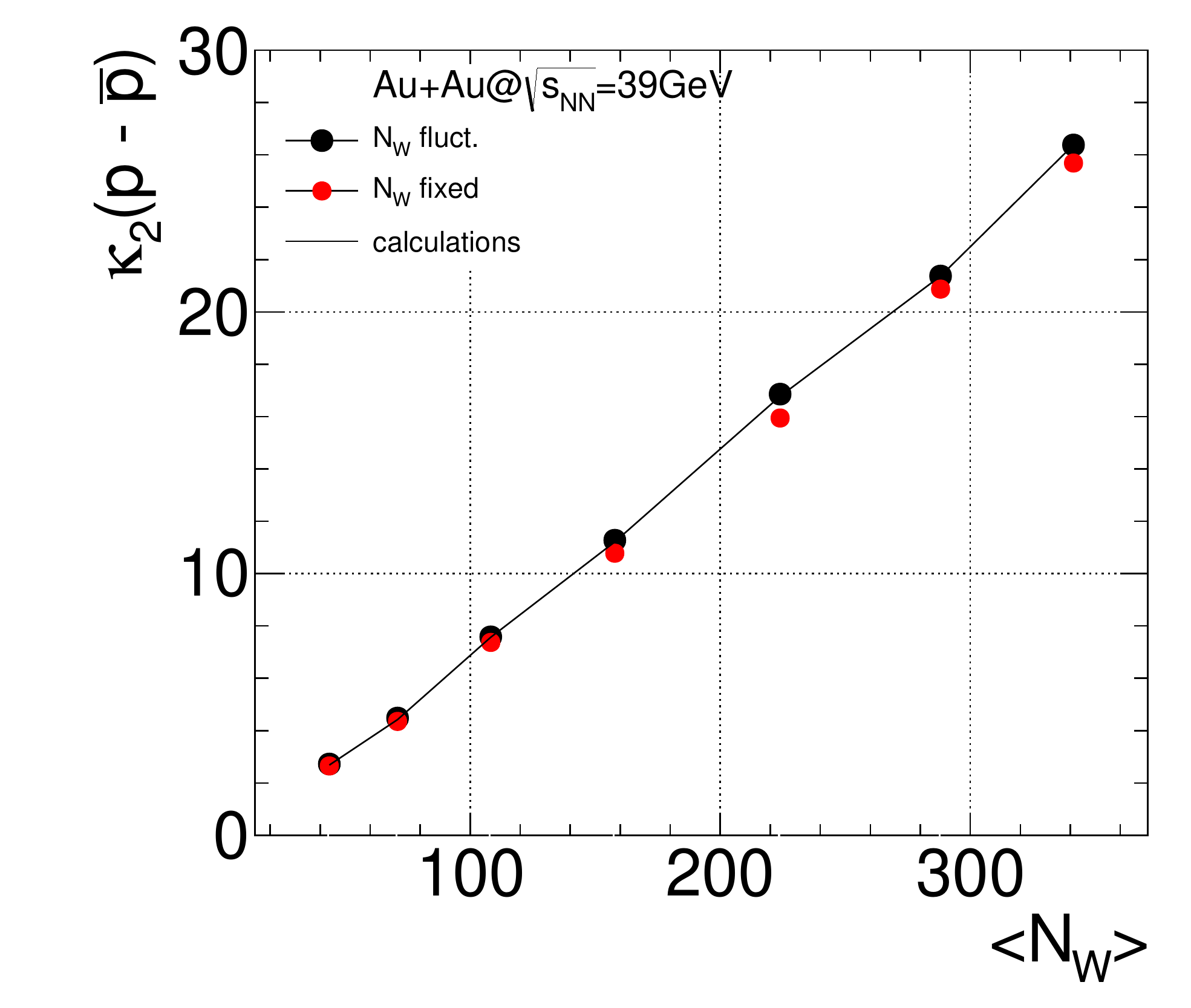}
 \includegraphics[width=0.45\linewidth,clip=true]{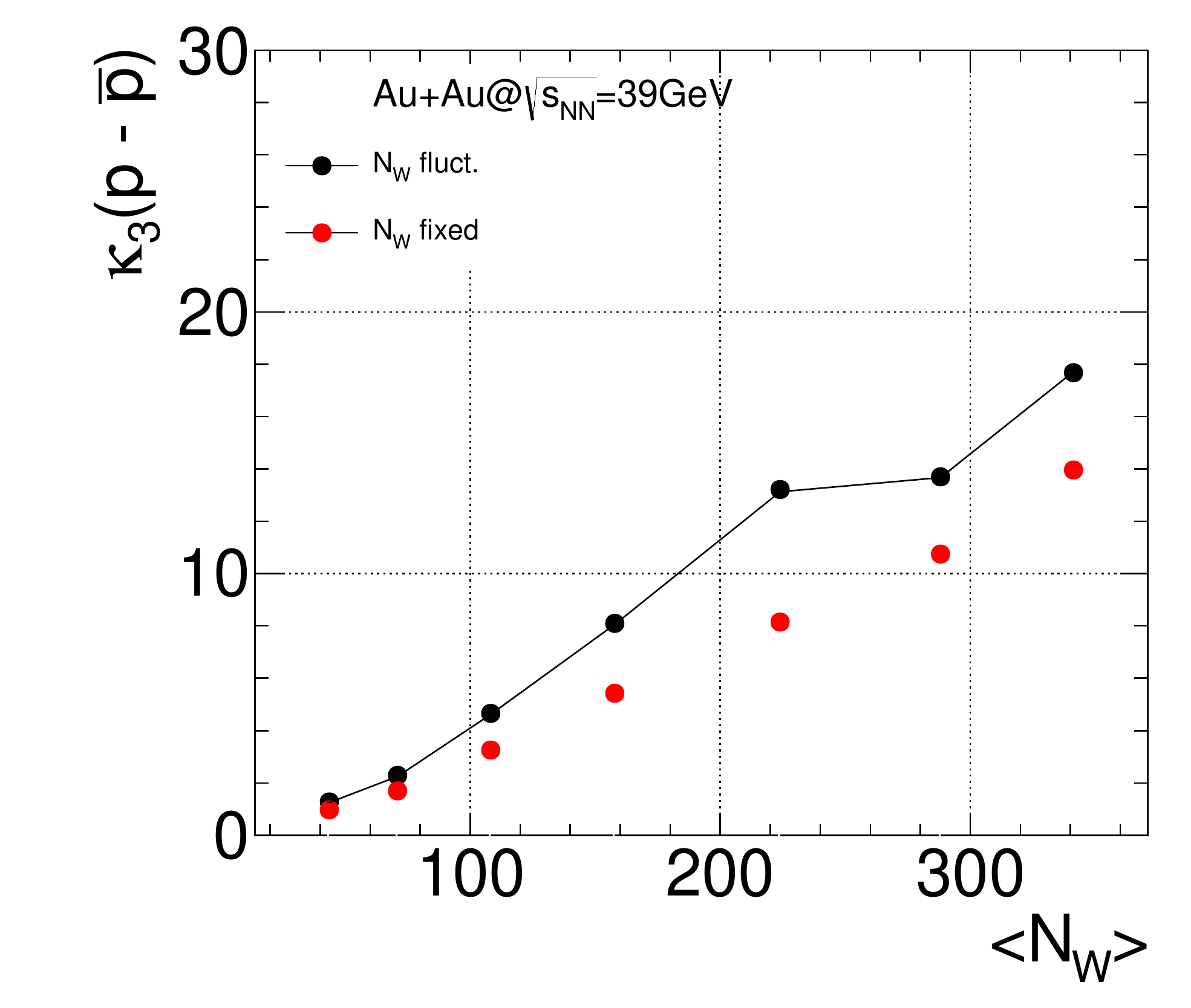}
 \caption{Second (left panel) and third (right panel) cumulants of
   net-protons for Au+Au collisions at $\sqrt{s_{NN}}$=39GeV. Red
   points correspond to keeping the number of wounded nucleons fixed,
   while for black points the fluctuations of wounded nucleons are
   included.The  black lines are calculated using eqs.~\ref{cum2} and~\ref{cum3}. }
\label{star391}
\end{figure}  

\begin{figure}[htb]
 \includegraphics[width=0.45\linewidth,clip=true]{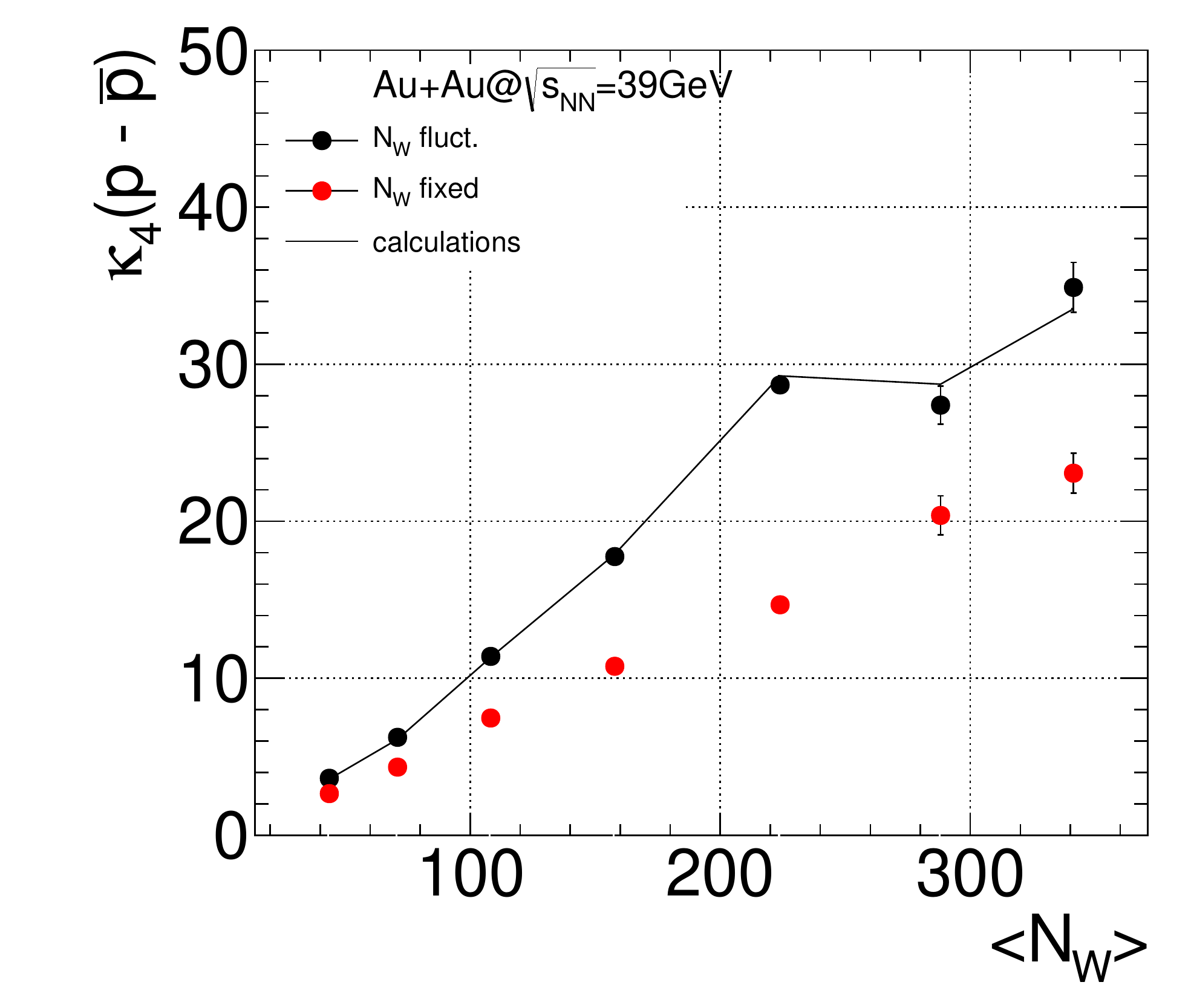}
 \includegraphics[width=0.45\linewidth,clip=true]{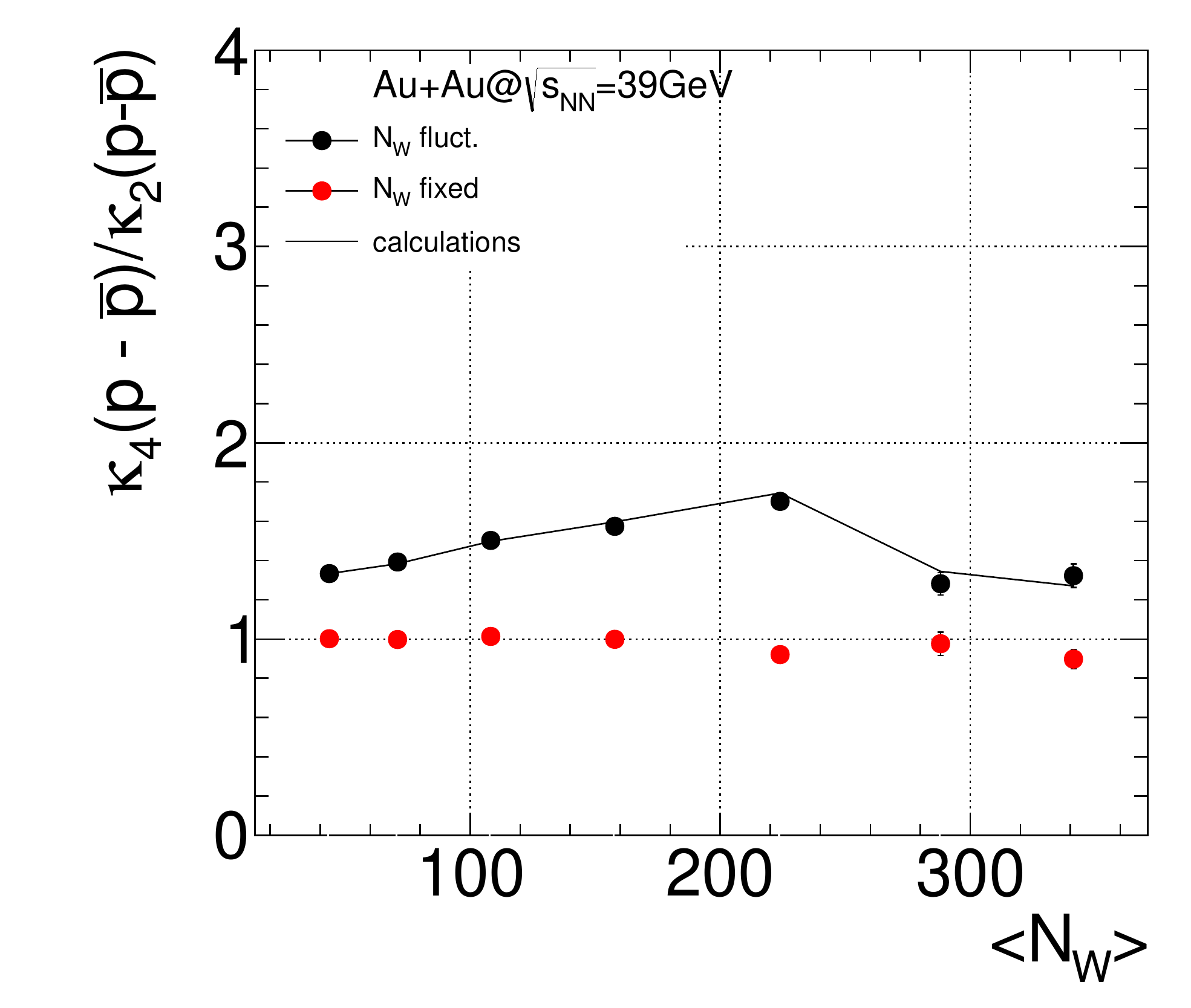}
 \caption{Left panel: Fourth cumulants of net-protons for Au+Au
   Collisions at $\sqrt{s_{NN}}=39GeV$. Right panel: Ratio of fourth
   and second cumulants. Red points correspond to fixed number of
   wounded nucleons while, for the black points, the fluctuations of
   wounded nucleons are included. The  black lines are calculated using eqs.~\ref{cum2} and~\ref{cum4}.}
\label{star392}
\end{figure}  

We observe that the CBWC reduces the overall level of fluctuations
significantly but cannot fully eliminate the participant
fluctuations. This can also be seen in Fig.~\ref{cent1} from the
2-dimensional scatter plots there, where even for a fixed value of
charged particles the number of wounded nucleons still fluctuates. On
the other hand, the incoherent addition of data from intervals with
very small centrality bin width will likely distort the physics we are
after since the correction also eliminates true dynamical
fluctuations. The CBWC in particular reduces true dynamical
correlations.  This is particularly relevant for searches for a
critical endpoint in the phase diagram. One of the signatures of such
a critical endpoint is that near it the dynamical correlation length
will increase rapidly (see above). Since particle production is also
likely to be affected, the sensitivity of a search for a critical
endpoint will be diminished if too small centrality bins are used.

In Fig.~\ref{star4} we show the results for participant fluctuations
for the fourth cumulants of net-protons and their ratio to the second
cumulants. Even for very fine centrality bin widths we observe up to
$40\%$ deviations from the baseline.  Furthermore, participant
fluctuations are suppressed less than shown in Figs.~\ref{star23}
and~\ref{star4} if autocorrelations with the charged particles used
for the centrality determination are not removed entirely. We note, in
this context, that a significant contribution to net-proton
fluctuations will originate from fluctuations of the number of net
$\Delta$ baryons. This will introduce strong pion-proton correlations
into the sample implying that a part of the auto-correlation problem
survives, even if one excludes protons and antiprotons from the data
used for centrality determination.

Like in case of protons at $\sqrt{s_{NN}}=2.76$ TeV (see the left
panel of Fig.~\ref{k4}), we observe small effects of the participant
fluctuations for the most central bin in Fig.~\ref{star4}.  As
explained above, this stems from the negative values of
$\kappa_{3}(N_{W})$ and $\kappa_{4}(N_{W})$. However, this also
depends on the mean number of particles or net-particles.  To show
this explicitly we present, in Figs.~\ref{star391} and~\ref{star392},
cumulants of net-protons for Au+Au collisions at $\sqrt{s_{NN}}$=39
GeV. Mean values of protons and antiprotons are taken
from~\cite{STARDATA}.  For the second cumulants of net-protons we
observe quite small contributions from participant
fluctuations. However, for the third and fourth cumulants these
contributions are significant.  Moreover, even for the most central
bin $\kappa{4}(p-\bar{p})/\kappa{2}(p-\bar{p})$ deviates from unity if
participant fluctuations are included.

\section{Global conservation laws}
\label{conserv}
In this section we will demonstrate a procedure for selecting an
"optimized" acceptance for fluctuation analysis. For clarity we will
focus on net-baryon fluctuations, though our approach is valid for any
conserved charges. We remind at this point that critical net-particle
fluctuations are predicted within the Grand Canonical Ensemble (GCE)
formulation of thermodynamics. In this formulation, the net-baryon
number is not conserved in each micro-state, hence it fluctuates.  In
order to make sense of that, chemical potentials are introduced which
guaranties net-baryon number conservation on the average. In order to
compare theoretical calculations within GCE, such as HRG model and
LQCD, to experimental results, the requirements of GCE have to be
achieved in experiments. This is typically done by analysing the
experimental data in the finite acceptance by imposing cuts on
rapidity and/or mean transverse momentum of detected
particles. However, if the selected acceptance window is too small,
the possible dynamical correlations we are after will also be strongly
reduced~\cite{KochFluct} and consequently, net baryons will be
distributed according to the difference of two independent Poisson
distributions, the Skellam distribution. This statement is
analytically proven below. On the other hand, by enlarging the
acceptance, in order to catch dynamical fluctuations, correlations due
to baryon number conservation will be significant.  The aim of this
section is to estimate the contribution from the conservation laws and
subtract it from the measured fluctuation signals.

In order to get a quantitative estimate for what means "large"
acceptance we will model the finite acceptance with the binomial
distribution.

We first define the acceptance factor for baryons as the ratio of mean number
of detected baryons $\left< N_{B}^{acc}\right>$ to the number of
baryons in the full phase space $\left< N_{B}^{4\pi}\right>$:

\begin{equation}
 \alpha = \frac{\left< N_{B}^{acc} \right>}{\left< N_{B}^{4\pi}\right>}.
\label{accDef}
\end{equation}
Furthermore, we assume the same acceptance factor for anti-baryons.
Given the number of baryons $N_{B}$ in the full phase space, the
probability of measuring $n_{B}$ baryons in the acceptance is

\begin{equation}
B\left(n_{B};N_{B},\alpha \right) = \frac{N_{B}!}{n_{B}!\left(N_{B}-n_{B}\right)!}\alpha^{n_{B}}\left(1-\alpha\right)^{N_{B}-n_{B}},
\label{binomial11}
\end{equation}

If the number of baryons in $4\pi$ are distributed according to some
probability distribution $P(N_{B})$ the corresponding multiplicity
distribution in the acceptance will then be

\begin{equation}
P(n_{B})=\sum_{N_{B}}B(n_{B};N_{B},\alpha)P(N_{B}).
\label{binomial12}
\end{equation}

The moments of the measured baryon distributions can be then  calculated

\begin{equation}
\left<n_{B}\right> = \sum_{n_{B}=0}^{\infty}n_{B}P(n_{B})=\alpha\left<N_{B}\right>,
\label{binomial1}
\end{equation}

\begin{equation}
\left<n_{B}^{2}\right> =
\sum_{n_{B}=0}^{\infty}n_{B}^{2}P(n_{B})=\alpha^{2}\left<N_{B}^{2}\right> + \alpha(1-\alpha)\left<N_{B}\right>.
\label{binomial2}
\end{equation}

In a similar way corresponding moments for the anti-baryons can be derived:

\begin{equation}
\left<n_{\bar{B}}\right> = \sum_{n_{\bar{B}}=0}^{\infty}n_{\bar{B}}P(n_{\bar{B}})=\alpha\left<N_{\bar{B}}\right> ,
\label{binomial3}
\end{equation}

\begin{equation}
\left<n_{\bar{B}}^{2}\right> =
\sum_{n_{\bar{B}}=0}^{\infty}n_{\bar{B}}^{2}P(n_{\bar{B}})=\alpha^{2}\left<N_{\bar{B}}^{2}\right> + \alpha(1-\alpha)\left<N_{\bar{B}}\right>.
\label{binomial4}
\end{equation}

Finally, the mixed moment of baryons and anti-baryons are obtained 

\begin{equation}
\left<nn_{\bar{B}}\right> = \alpha^{2}\left<N_{B}N_{\bar{B}}\right>.
\label{binomial5}
\end{equation}

The second cumulant of net baryons inside the acceptance can be written as

\begin{equation}
\kappa_{2}\left({n_{B}-n_{\bar{B}}}\right) = \kappa_{2}\left(n_{B}\right) + \kappa_{2}\left(n_{\bar{B}}\right) - 2\left(\left<n_{B}n_{\bar{B}}\right> - \left<n_{B}\right>\left<n_{\bar{B}}\right>\right).
\label{secondCumAcceptance}
\end{equation}

Using eqs.~\ref{binomial1}-\ref{binomial5} in eq.~\ref{secondCumAcceptance} we obtain

\begin{equation}
\frac{\kappa_{2}\left(n_{B}-n_{\bar{B}}\right)}{\kappa_{2}\left(Skellam\right)}=\frac{\kappa_{2}\left(n_{B}-n_{\bar{B}}\right)}{\alpha\left(\left<N_{B}\right>+\left<N_{\bar{B}}\right>\right)}=\alpha\frac{\kappa_{2}\left(N_{B}-N_{\bar{B}}\right)}{\left<N_{B}\right>+\left<N_{\bar{B}}\right>}+1-\alpha,
\label{conserv1}
\end{equation}
here $\kappa_{2}\left(Skellam\right)$ refers to the second cumulant of
the Skellam distribution, which, according to
eq.~\ref{Skellam_definition} is equal to $\left<n_{B} +
n_{\bar{B}}\right>$.

The eq.~\ref{conserv1} leads to

\begin{equation}
\frac{\kappa_{2}\left(n_{B}-n_{\bar{B}}\right)}{\kappa_{2}\left(Skellam\right)}=1-\alpha.
\label{conserv2}
\end{equation}
because net-baryons do not fluctuate in $4\pi$, \emph{i.e},
$\kappa_{2}\left(N_{B}-N_{\bar{B}}\right)$ in eq.~\ref{conserv1}
vanishes.

Eq.~\ref{conserv2} shows that fluctuations of net-baryons inside the
acceptance will be modified due to the baryon number
conservation. Moreover, the modification depends only on the
acceptance factor $\alpha$, defined in eq.~\ref{accDef}.  We first
examine a number of useful properties of eqs.~\ref{conserv1}
and~\ref{conserv2}. When $\alpha$ approaches zero the
eq.~\ref{conserv2} converges to unity. This means that, in a small
acceptance, net-baryon distributions can be described with the Skellam
probabilities. On the other hand, with increasing $\alpha$, the
fluctuations of net-baryons decrease because of the increasingly
significant effect of overall baryon number conservation, and
eventually vanish when $\alpha$ becomes 1. Moreover, from
eq.~\ref{conserv1} it is evident that, if in a larger acceptance the
net-baryon fluctuations follow the Skellam distribution, then in any
smaller acceptance the multiplicities will also be distributed
according to the Skellam distribution. Indeed, in this case
$\kappa_{2}\left(N_{B}-N_{\bar{B}}\right)/\left<N_{B} +
N_{\bar{B}}\right> = 1$ which leads to
$\kappa_{2}\left(n_{B}-n_{\bar{B}}\right)/\left<n_{B} +
n_{\bar{B}}\right> = 1$. The latter is important and once again
underlines the importance of large acceptance.

Experiments typically report on net-proton cumulants which are used as
a proxy for the net-baryons. The validity of this assumption is
fulfilled at the LHC energies~\cite{Kitazawa}. In order to correct net proton
distributions for the baryon number conservation eq.~\ref{conserv2}
can still be employed by redefining the the $\alpha$ (see
eq.~\ref{accDef}) parameter as

\begin{equation}
 \alpha = \frac{\left< N_{p}^{acc} \right>}{\left< N_{B}^{4\pi}\right>},
\label{accDef2}
\end{equation}

where, $\left<N_{p}^{acc}\right>$ refers to the mean number of protons
inside the acceptance.

\section {Conclusion}

In summary, we studied non-dynamical contributions to fluctuations of
net-protons within the Wounded Nucleon Model. Since the impact
parameter is not measurable in a nuclear collision such fluctuations
of participants cannot be avoided experimentally. To study their
impact, we developed a Monte Carlo method to describe realistically
the centrality dependence of the collision geometry and its influence
on higher moments of net-baryon distributions. To this end we provide
analytic relations between net-particle and wounded nucleon cumulants
for any cumulant order.  Furthermore, we discuss a procedure
for selecting an 'optimal' acceptance for fluctuation measurements.

The results of our studies exhibit a strong centrality and energy dependence for
these non-dynamical fluctuations. The magnitude of the effect is very
significant, exceeding the fluctuations computed for a non-interacting
hadron resonance gas by more than a factor of 2.  Only for the case of
vanishing mean values of net-protons, as observed at mid-rapidity for
LHC energies, their second and third cumulants are found to be
independent of participant fluctuations, while higher even moments are
significantly affected. At lower energies, where the mean numbers of
protons and antiprotons are different, the second and all higher
cumulants depend strongly on fluctuations of wounded nucleons (or
participants).
 The simplest way to use our approach in the interpretation of
experimental data is to start from a particular theoretical prediction
and fold the results with our calculations for fluctuations of
participants and conservation of net-baryon number before any detailed
comparison with experiment. A more ambitious and ultimately necessary
program would be to unfold experimental data in the inverse
approach. This will require a very detailed understanding of
experimental resolutions and efficiencies. The effects we have
observed, in particular for higher than 2nd cumulants, are so
significant at all beam energies that their inclusion into analysis
procedures seem mandatory before quantitative physics conclusions from
fluctuation data can be obtained.

\section{ACKNOWLEDGMENTS}
This work is part of and supported by the DFG Collaborative Research
Centre "SFB 1225 (ISOQUANT)". The authors acknowledge stimulating
discussions with Bengt Friman, Volker Koch, and Krzysztof Redlich. One
of us (pbm) would like to thank Jochen Thaeder for insightful remarks
concerning the intricacies of fluctuation analyses.

\end{document}